\documentclass[twocolumn,trackchanges]{aastex701}
\usepackage{amsmath,amssymb}
\usepackage{hyperref}
\usepackage{cleveref}
\usepackage{subcaption}
\usepackage{graphicx}
\graphicspath{{./}{figures/}}
\usepackage{placeins}

\newcommand{\figref}[1]{\hyperref[#1]{Figure~\ref*{#1}}}

\newcommand{\Hb}{H$\beta$}

\newcommand{\Lmon}{$L_{\rm 5100}$}
\newcommand{\mbh}{$M_{\rm BH}$}

\newcommand{\RBLR}{$R_{\rm BLR}$}

\newcommand{\Mobs}{$M_{\rm obs}$}

\newcommand{\Eobs}{$\lambda_{\rm obs}$}
\newcommand{\Tobs}{$\tau_{\rm obs}$}
\newcommand{\Tmodel}{$\tau_{\rm model}$}

\begin{document}
\title{The Broad-Line Region Size–Luminosity Relation in AGN Constrained by the Viewing Angle with the FRADO Model}
%\title{Exploring the Broad-Line Region Size–Luminosity Relation in AGN using the FRADO Model with Viewing Angle Constraints}

\author{Junhee Myong}
\affiliation{Department of Physics \& Astronomy, Seoul National University, Seoul 08826, Republic of Korea, jhwoo@snu.ac.kr}
\email{r136a136@gmail.com}

\author[orcid=0000-0002-8055-5465,gname='Jong-Hak', sname='Woo']{Jong-Hak Woo} 
\affiliation{Department of Physics \& Astronomy, Seoul National University, Seoul 08826, Republic of Korea, jhwoo@snu.ac.kr}
\affiliation{SNU Astronomy Research Center, Seoul National University, Seoul 08826, Republic of Korea}
\email{jhwoo@snu.ac.kr}

\author[orcid=0000-0002-7604-9594,gname='Mohammad Hassan', sname='Naddaf']{Mohammad Hassan Naddaf}
\affiliation{Institut d’Astrophysique et de Géophysique, Université de Liège, Allée du six août 19c, B-4000 Liège (Sart-Tilman), Belgium}
\email{mh.naddaf@uliege.be}  

\author[orcid=0000-0001-5848-4333, gname='Bo\.zena',sname='Czerny']{Bo\.zena Czerny}
\affiliation{Center for Theoretical Physics, Polish Academy of Sciences, Al. Lotników 32/46, 02-668 Warsaw, Poland}
\email{bcz@cft.edu.pl}

\author{Jimin Kim}
\affiliation{Department of Physics \& Astronomy, Seoul National University, Seoul 08826, Republic of Korea, jhwoo@snu.ac.kr}
\email{paul5282@snu.ac.kr}

\begin{abstract}
We investigate the H$\beta$ broad-line region (BLR) size--luminosity (\RBLR-\Lmon) relation of active galactic nuclei (AGN) using a Bayesian viewing angle inference framework based on the Failed Radiatively Accelerated Dusty Outflow (FRADO) model.
First, we update the FRADO model by introducing a phase factor to account for the anisotropic visibility of the illuminated cloud surface.
Second, we construct a Bayesian framework to constrain the viewing angle by matching the FRADO-based \RBLR\ to the measured H$\beta$ lags of 151 AGN from \citet{Wang24}. 
We successfully reproduce the slope and intrinsic scatter of the observed \RBLR-\Lmon\ relation, finding that the inferred viewing angles are broadly consistent with the estimates from dynamical modeling of velocity-resolved reverberation data, albeit with substantial scatter. As the observed \Hb\ time lag and line velocity depend on the viewing angle, the black hole mass (\mbh) calibrated from our work is typically larger than the reverberation-mapping estimates using an average virial factor.
Our results provide a physically motivated interpretation of the \RBLR-\Lmon\ relation and the systematic uncertainties in reverberation and single-epoch \mbh\ estimates.
\end{abstract}
\keywords{Active galactic nuclei, Broad-line region, Reverberation mapping, Supermassive black holes, Active galaxies}

\section{Introduction}

Supermassive black holes (SMBHs) at the centers of galaxies are known to play a significant role in galaxy evolution through feedback during the active galactic nucleus (AGN) phases
\citep[e.g.,][]{
%McNamara07, Cattaneo09, 
Fabian12, Heckman14, 
%Harrison17, Morganti17, 
Hardcastle20, Harrison24}. 
Understanding AGN feedback and black hole physics requires reliable black hole mass (\mbh) measurements.
Reverberation mapping (RM) provides \mbh\ for type-1 AGN based on the time lag between continuum and broad emission-line variability 
\citep{Blandford82, Peterson93}, constraining a characteristic size of the broad-line region (\RBLR).

One of the key results from the RM studies is the empirical relation between \RBLR\ and continuum luminosity at 5100\AA\ (\Lmon), 
%which relates \RBLR\ to the monochromatic continuum luminosity, \Lmon\
which enables indirect \mbh\ estimates by inferring \RBLR\ from \Lmon\ using a single-epoch spectrum
\citep[e.g.,][]{Kaspi00, WooUrry02, Peterson04, 
%GreeneHo05, 
Vestergaard06, Bentz09}.
%Using 41 local AGN and careful starlight decomposition, 
For example, \citet{Bentz13} reported a tight \RBLR--\Lmon\ relation with a slope close to 0.5, which is 
%and an intrinsic scatter of 0.19 dex, 
consistent with simple photoionization models. This relation has been widely used for single-epoch \mbh\ estimates \citep[e.g.,][]{Shen11,Peterson14,Baron19,Rakshit20,Wushen22,Pan25,Pucha26}.
% \citealt{Wang24}). 
%Furthermore, the \RBLR--\Lmon\ relation has a potential as a standard candel, since the \RBLR\ measurements from RM can predict the optical luminosity.
%physical size and luminosity of distant objects 
%(\citealt{Watson18}; \citealt{Czerny13}; \citealt{Melia15}; \citealt{Martinez19}; \citealt{Jaiswal25}).

However, later studies showed that super-Eddington AGN tend to deviate from the \RBLR-\Lmon\ relation
 \citep[e.g.,][]{Du16, Du18, Martinez20, Li21,Woo24}, although the physical reason is not yet clear \citep[see discussion by][]{Woo24}. 
These results have motivated calibrations  accounting for the Eddington ratio effect \citep[see][]{Du18,Du19,Yu19,Woo26}. 
By uniformly analyzing \Hb\ lags for a large sample of AGN, \citet{Wang24} recently updated the \RBLR--\Lmon\ relation,
reporting that the slope of the relation is $\sim$0.4 with an intrinsic scatter of $\sim$0.2 dex \citep[see also][]{Woo24}. 
\citet{Woo26} showed that three parameters, \RBLR, \Lmon, and gas velocity, define a tight fundamental plane, suggesting the fundamental plane relation is more suitable for single-epoch mass estimates than the two-parameter \RBLR-\Lmon\ relation. 

Several theoretical models have been proposed to explain the origin of the BLR
(e.g., \citealp{Ferland89,Emmering92,Murray95,Czerny11,Begelman17,Baskin18,Hopkins25})
and the \RBLR--\Lmon\ relation
\citep[see also][]{Netzer93,Murray95,Czerny11,Baskin18}. Among these models, the Failed Radiatively Accelerated Dusty Outflow (FRADO) model, first introduced by \citet{Czerny11}, describes the BLR as a population of \Hb-emitting clouds lifted above the accretion disk by radiation pressure acting on dust, while gravity from the central black hole eventually causes the clouds to fall back after dust sublimation. The FRADO model predicts the inner and outer boundaries of the BLR, asymmetric emission-line profiles (\citealt{Naddaf24}), and the dependence of the virial factor on black hole mass (\citealt{Czerny17}).

The 2.5D FRADO model by \citet{Naddaf21} reproduces an \RBLR--\Lmon\ relation similar to that reported by \citet{Bentz13} based on the particle-based three-dimensional implementation \citep[see also][]{Naddaf25, Jaiswal25}. 
However, the 2.5D FRADO model
% of \citet{Naddaf25} 
does not explicitly include anisotropic visibility of the illuminated cloud surface to the line-of-sight (LOS), although such anisotropy has been suggested by BLR dynamical modeling studies (\citealt{Pancoast12}; \citealt{Pancoast14}; \citealt{Williams18}). Since the far-side clouds may contribute more than near-side clouds to the 
transfer function used to define the representative time lag, 
%but no physical weighting scheme was applied to the transfer function to enhance the relative contribution of the far-side response. 
a phase-dependent weighting scheme is needed to properly calculate the \Hb\ time lag. 
%self-consistently, rather than selecting among multiple lag definitions.

%In addition to the intrinsic cloud distribution and anisotropic emission response, 
The viewing angle is another key factor in determining the LOS-dependent time lag. In practice, however, the viewing angle remains difficult to determine observationally for individual AGN, while several recent studies have attempted to constrain it using dynamical modeling of velocity-resolved RM data (e.g., \citealt{Pancoast11,Pancoast12,Williams18,Wang26}) or interferometric phase spectra (e.g., \citealt{Sturm18}; \citealt{Amorim20}; \citealt{Amorim21}; \citealt{GRAVITY24}). 

With these motivations, we investigate whether the FRADO model, combined with a viewing angle inference framework, can reproduce the observed \RBLR--\Lmon\ relation, by determining the representative time lag from the transfer function after correcting for the anisotropic visibility of each cloud. 
In our Bayesian framework, we constrain the viewing angle ($\theta$), which determines the projection effects on the observed time lag, line velocity, and \mbh.
% estimates of AGN properties, $M$ and $\lambda$.
%The structure of this paper is as follows. 
In Section~\ref{sec:lagModeling}, we present the update of the 2.5D FRADO model.
%and the updated method for calculating the time lag from the 3D particle distribution. 
The Bayesian framework for the viewing angle calibration is presented in Section~\ref{sec:bayesian}. In Section~\ref{sec:results}, we present the Bayesian inference results, including the inferred viewing angle and AGN properties.
% and examine how well our model reproduces the observations. 
Discussion is presented in Section~\ref{sec:discussion}, followed by summary and conclusions in Section~\ref{sec:sum&conclusion}.
% we summarize our main conclusions.

\section{Time Lag constrained by the FRADO Model}
\label{sec:lagModeling}
In this section, we describe the update of the FRADO model and the calculation of \Hb\ time lag. First, we describe the implementation of the BLR in the FRADO model (\S~2.1). Second, we present the phase factor to include anisotropic visibility of BLR clouds (\S~2.2). Third, we describe the grid model used to calculate time lag over a continuous parameter space (\S~2.3). 

\subsection{Calculation of Transfer Function}
\label{sec:tau_cloud}

\begin{figure*}[ht!]
  \centering
  \plotone{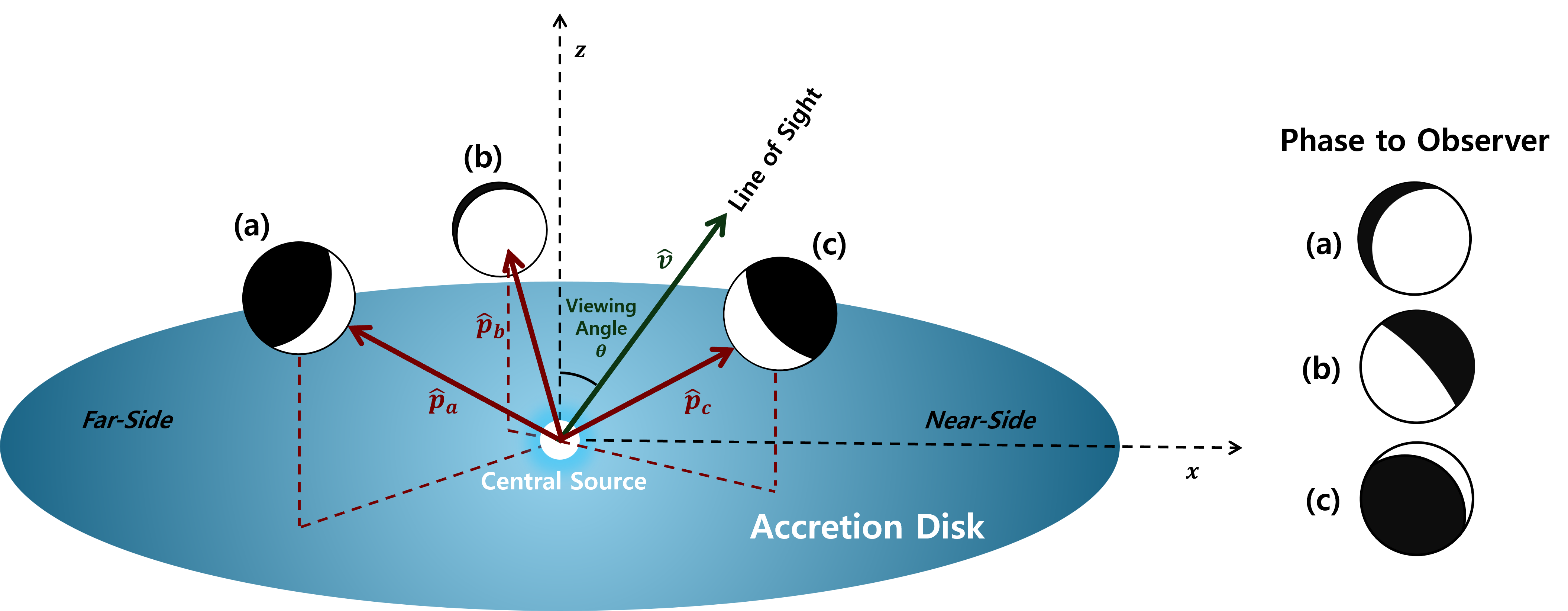}
  \caption{Illustration of the phase factor (Equation~\ref{eq:phase}). The position vectors of three arbitrary clouds are denoted with red arrows. For each cloud, the illuminated surface due to the central source is shown in white, whereas the right panel shows the reflected surfaces to the LOS in white.}
  \label{Fig 1}
\end{figure*}

The BLR described by the FRADO model is an ensemble of non-interacting clouds, whose time-independent particle distribution is used to calculate the representative H$\beta$ time lag ($\tau_{model}$). The particle distribution is determined by the black hole mass and luminosity (or equivalently, Eddington ratio $\lambda$), which set the gravitational potential of the central black hole and radiation pressure from the accretion disk, respectively. Once lifted from the disk surface, the clouds evolve under the combined effects of gravity and radiation pressure. The disk region contributing to the radiation pressure on each cloud is determined using the simplified shielding model introduced in \citet{Naddaf20} and \citet{Naddaf21}. When a cloud reaches the dust sublimation temperature, its dust is destroyed, radiation pressure ceases, and the cloud falls back toward the disk. Together with gravitational rotation, this vertical oscillatory motion produces the characteristic three-dimensional distribution of BLR clouds.

For a given BLR geometry specified by black hole mass and $\lambda$, the viewing angle ($\theta$) determines the LOS-dependent $\tau_{model}$, which is calculated in three steps: computing the time lags of individual clouds, constructing the phase-weighted transfer function and deriving $\tau_{model}$, and extending the grid model to a continuous parameter space.

We start by calculating the time lag of individual clouds. The time lag of each cloud is defined as the light-travel path difference between the reprocessed line-emission and the continuum. Assuming that the continuum propagates directly from the central source to the observer, the time lag of each cloud is given by
\begin{equation}
\tau_{cloud} = \frac{\left\lVert \mathbf{p}_{cloud} \right\rVert - \mathbf{p}_{cloud} \cdot \hat{\mathbf{v}}}{c},
\end{equation}
where 
%$\tau_{cloud}$ is the time lag of an individual cloud, 
$\mathbf{p}_{cloud}$ is the cloud position vector relative to the central black hole, $\hat{\mathbf{v}}$ is the LOS unit vector, and c is the speed of light.

Then, we construct the transfer function by collecting the time lags of individual clouds. 
%A representative lag is then derived from the transfer function and adopted as the model time lag, $\tau_{model}$.

\subsection{Introduction of a Phase Factor to the Transfer Function}
\label{sec:phase_factor}

We introduce a phase factor into the transfer function to account for the anisotropic visibility of the illuminated side of each cloud. Following \citet{Naddaf25}, we first assign each cloud an H$\beta$ emissivity determined by the incident photon flux. Then we multiply the emissivity by a dimensionless phase factor that accounts for the anisotropic visibility of the illuminated cloud surface. Because BLR clouds are expected to be optically thick to the central ionizing radiation, only the surface facing the central source is illuminated, and only part of this surface is visible to the observer (see \figref{Fig 1}). 
%The final weight assigned to each cloud is given by the product of the emissivity weight and the phase factor. 

\begin{figure}[!t]
  \centering
    \vspace{0pt}
    \centering
    \includegraphics[height=0.43\textwidth]{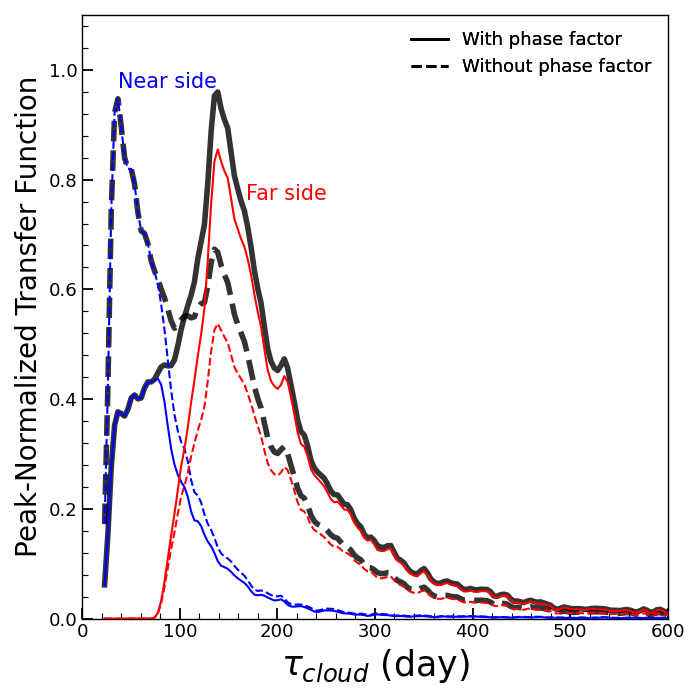}
    \caption{
    Example transfer functions of an AGN with \mbh= 10$^{9.2}$M$_{\odot}$ and $\lambda$=0.16. The total transfer function (black line) is calculated using the \Hb\ emissivity with and without the phase factor (dashed and solid lines, respectively). The corresponding far-side and near-side transfer functions are shown in red and blue, respectively.}
  \label{tf}
\end{figure}

Assuming parallel incident rays along the $\hat{\mathbf{p}}_{cloud}$ direction and spherical clouds, we define the phase factor as the fraction of the projected cloud surface visible to the observer that is illuminated by the central source as:
\begin{equation}
Phase \equiv \frac{1 - \hat{\mathbf{p}}_{cloud} \cdot \hat{\mathbf{v}}}{2},
\label{eq:phase}
\end{equation}
where $\hat{\mathbf{p}}_{cloud}$ is the normalized position vector of a cloud. 
Although FRADO clouds are point particles, we introduce a spherical shape to define the dimensionless phase factor.
% which is independent of cloud size. 
The phase factor is unity when an observer sees the fully illuminated surface of a cloud, and zero when the illuminated surface is invisible along the LOS. As a result, clouds located on the far-side of the BLR from an observer have larger phase factors than those on the near-side, increasing the contribution of the far-side BLR to the transfer function \citep[see also ][]{Pancoast12} as illustrated in \figref{tf}. 
%Examples of the resulting transfer functions, their near-side and far-side components, and their relative contributions are shown in 
%The phase factor suppresses the near-side response and enhances the far-side response. 
This effect of the phase factor becomes more pronounced with a larger $\theta$, as the projected separation of the two sides increases and the median of the 
transfer function shifts toward longer lags.

\begin{figure}[!t]
  \centering
    \vspace{0pt}
    \centering
    \includegraphics[height=0.46\textwidth]{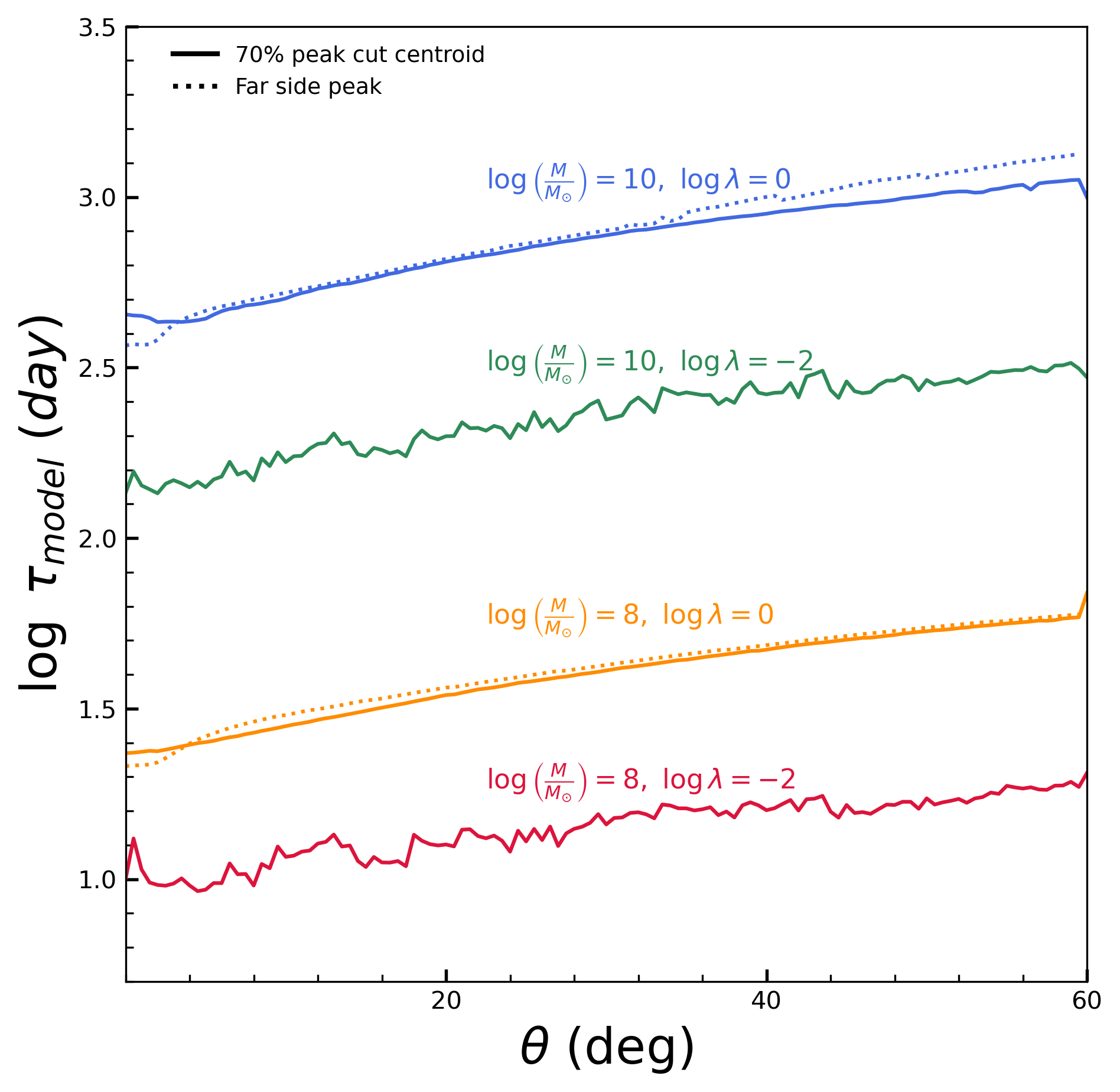}
    \caption{Viewing angle dependence of $\tau_{model}$ for AGN with various $M$ and $\lambda$. $\tau_{model}$ is defined as the centroid above the 70$\%$ peak threshold (solid) or the far-side peak (dotted lines).}
    %which is nearly identical to the 70$\%$ peak threshold definition.}
  \label{tau_model}
\end{figure}

We define $\tau_{model}$ as the centroid of the transfer function region above 70$\%$ of the peak. This definition yields time lags broadly consistent with the Gaussian-smoothed far-side peak adopted by \citet{Naddaf25}. Although $\tau_{model}$ differs from the cross-correlation function (CCF) lags commonly measured above 80$\%$ of the CCF peak in RM studies, it provides a similar characteristic BLR response time.

We present $\tau_{model}$ as a function of viewing angle for various pairs of $M$ and $\lambda$ in \figref{tau_model}, showing that $\tau_{model}$ generally increases with viewing angle. In the following analysis, we will use the viewing angle dependent $\tau_{model}$ as a representative time lag of the BLR.

%Examples of $\tau_{model}$ obtained for different combinations of $M$, $\lambda$, and $\theta$ and for alternative time lag definitions are shown in 

%Larger values of $M$ and $\lambda$ generally correspond to higher AGN luminosities and larger characteristic BLR sizes, yielding longer $\tau_{model}$. At fixed $\theta$, the updated FRADO model produces an \RBLR--\Lmon\ slope similar to those of the \citet{Bentz13} relation and analytical FRADO (\citealt{Czerny15}), consistent with the far-side peak results from \citet{Naddaf25}. At fixed $M$ and $\lambda$, $\theta$ introduces an additional dependence in $\tau_{model}$ because the phase-weighting scheme increases the relative contribution of far-side clouds, whose $\tau_{cloud}$ values become longer at larger $\theta$. Over the range considered in this work, the viewing angle dependence changes $\tau_{model}$ by up to $\sim0.5$ dex.

\subsection{Extension to a Continuous Parameter Space}

We extend the discrete FRADO lag grid to a continuous parameter space to evaluate $\tau_{model}$ for arbitrary input parameters of mass and Eddington ratio. The cloud distributions are computed on a two-dimensional grid spanning $\log(M/M_\odot)=6$--$10$ and $\log\lambda=-2$--$0$, with $\tau_{model}$ evaluated over $\theta=0^\circ$--$60^\circ$, appropriate for type 1 AGN. Rather than interpolating the full four-dimensional hypersurface $(\log (M/M_{\odot}),\log \lambda,\theta,\log \tau_{model})$, we evaluate the grid at $0.5^\circ$ intervals in $\theta$ and perform interpolation or extrapolation in the $(\log (M/M_{\odot}),\log \lambda,\log \tau_{model})$ space at each fixed $\theta$. In the low-luminosity regime, some combinations of $M$ and $\lambda$ fall below the incident ionizing photon-flux threshold adopted in the 2.5D FRADO model (\citealt{Naddaf25}), $10^{18}\,{ cm^{-2}\,s^{-1}}$. Consequently, the populated grid points form a triangular region in the $(M,\lambda)$ plane rather than a complete rectangle.

For a set of input parameters ($\log (M/M_{\odot})$, $\log \lambda$, $\theta$)  with sufficient surrounding grid support, we estimate $\tau_{model}$ using local interpolation. We first identify the neighboring grid points along the $\log(M/M_{\odot})$ and $\log\lambda$ directions and compute local linear estimates of $\log\tau_{model}$. The two local estimates are then combined using an inverse-variance-weighted average to obtain the interpolated lag and its uncertainty.

If the input lies outside the populated region or lacks sufficient local grid support, we use plane-based extrapolation. In this case, we fit a plane to all populated grid points in the $(\log(M/M_\odot),\log\lambda,\log\tau_{model})$ space and evaluate it at the input values of $\log (M/M_{\odot})$ and $\log \lambda$. This treatment is justified because the FRADO time-lag surface at fixed $\theta$ is nearly planar in the $(\log(M/M_\odot),\log\lambda,\log\tau_{model})$ space, with a typical rms scatter of $\sim0.03$ dex or smaller. This hybrid interpolation and extrapolation scheme provides $\tau_{model}$ throughout the continuous $(M,\lambda)$ parameter space, including regions without sufficient local grid support.

\section{Bayesian Framework for constraining the viewing angle}
\label{sec:bayesian}

\begin{figure*}[!t]
  \centering
  \includegraphics[height=0.45\textwidth]{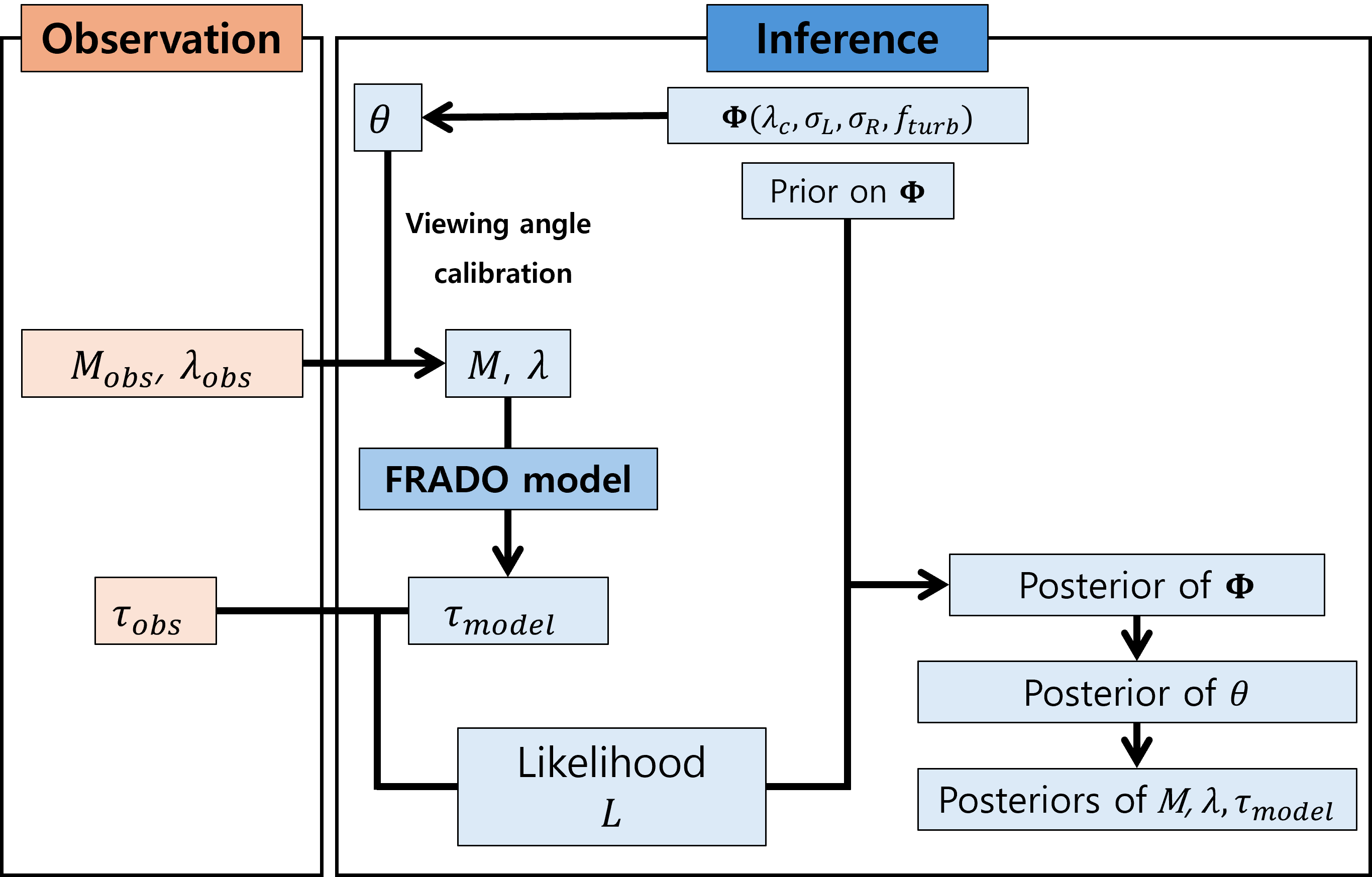}
  \caption{
  Bayesian framework used in this work. Given the observed quantities ($M_{obs}$, $\lambda_{obs}$, and $\tau_{obs}$), the framework uses the FRADO model and the viewing angle calibration of $M_{obs}$ and $\lambda_{obs}$ to infer $\theta$. Each trial parameter vector $\boldsymbol{\Phi}$ determines the relative probabilities of the trial viewing angles for each AGN. At each trial $\theta$, the corresponding $M$ and $\lambda$ are calibrated and used to compute $\tau_{model}$. The likelihood is evaluated by comparing $\tau_{model}$ with $\tau_{obs}$ and marginalizing over $\theta$. Markov chain Monte Carlo (MCMC) is then used to sample $\boldsymbol{\Phi}$. The posterior samples are propagated to infer the distributions of $\theta$, $M$, $\lambda$, and $\tau_{model}$.
}
  \label{flowchart}
\end{figure*}

In this section, we present a Bayesian framework for inferring the viewing angle $\theta$ by matching \Tmodel\ to the observed H$\beta$ lag (\Tobs), for a given AGN with the observed RM mass (\Mobs) and Eddington ratio (\Eobs). We apply this framework to a sample of RM AGN from \citet{Wang24}
in Section~\ref{sec:results}. 

%using a sample of AGN from \citet{Wang24},  for which RM mass (\Mobs) and Eddington ratio (\Eobs) are also available.
For each trial value of $\theta$ for an AGN with fixed \Mobs\ and \Eobs, we obtain $M$ and $\lambda$ as input parameters for the FRADO model and evaluate the likelihood of \Tobs\ given \Tmodel. Our Bayesian framework is summarized in \figref{flowchart}.
Note that $M_{obs}$ and $\lambda_{obs}$ represent the observed mass and Eddington ratio from the RM studies, while the unsubscripted quantities, $M$ and $\lambda$, denote the corresponding intrinsic parameters using the inferred $\theta$.
% for a given $\theta$ of \Mobs\ and \Eobs.

\subsection{$\theta$-Dependence of the Observed AGN Properties}

The observed RM \mbh\ is defined as
\begin{equation}
M_{obs} = \langle f\rangle c \tau_{obs}~ \Delta V_{los}^{2}/G,
\label{eq:mrm}
\end{equation}
where \Tobs\ is the \Hb\ lag, $\Delta V_{los}$ is the \Hb\ line width, and $\langle f\rangle$ is an average virial factor. 
%The viewing angle  dependence of $M_{obs}$ arises from both the time lag and the line width measurements. 
Since \Tobs\ changes as a function of the viewing angle to the LOS, and $\Delta V_{los}$ corresponds to the LOS velocity, 
$M_{obs}$ depends on $\theta$. 

The virial factor of individual AGN also depends on the viewing angle since it reflects the BLR geometry and kinematics projected along the LOS. In fact, the viewing angle is likely to be a dominant source of variation in the virial factor \citep[see also][]{Pancoast14,Grier17a,Williams18, Villafana26}. However, an average virial factor is often adopted in practice due to the difficulty of constraining a virial factor for individual AGN \citep[e.g.,][]{Woo15, Wang24}. This average virial factor may correspond to a representative viewing angle $\langle\theta\rangle$ of the adopted AGN sample, although the virial factor also depends on other BLR properties, such as the opening angle and radial extent of the BLR.
%In summary, $M_{obs}$ retains a viewing angle dependence because $\tau_{obs}$ and the line velocity vary with $\theta$, while a single average virial factor is applied to individual AGN.

The viewing angle dependence is also clear in the observed Eddington ratio defined as
\begin{equation}
\lambda_{\rm obs} \propto {10 \times L_{\rm 5100,obs} \over M_{\rm obs}}.
\label{eq:ER}
\end{equation}
The disk luminosity observed at 5100 Å ($L_{5100,obs}$) is often calculated from the measured flux assuming isotropic emission, although the disk emission is anisotropic with respect to the LOS. By adopting a simplified thin Lambertian disk prescription, we find that the observed continuum luminosity scales with $\cos\theta$ due to the projected emitting area.
However, the constant bolometric correction factor of 10 commonly applied to $L_{5100,obs}$ is calibrated for an AGN population and therefore implicitly corresponds to $\langle\theta\rangle$. Thus, the bolometric luminosity and Eddington ratio of each AGN need to be calibrated using its individual viewing angle.

\subsection{Constraining Representative Viewing Angle}
\label{theta_avg}

We quantify the viewing angle effect on the observed parameters by parameterizing the difference between the inferred $\theta$
and a representative viewing angle $\langle\theta\rangle$ under the assumption that only the viewing angle determines the virial factor.
We determine the representative viewing angle $\langle\theta\rangle$ corresponding to the average virial factor $\langle f\rangle$ in Equation~\ref{eq:mrm}. While $\langle f\rangle$ is typically adopted from the calibration of the \mbh–stellar velocity dispersion relation by \citet{Woo15}, individual estimates of $f$ and $\theta$ are not available for the AGN used in the calibration. Instead, we use 37 velocity-resolved RM AGN from \citet{Villafana26}, for which $f$ is calculated as the ratio of the \mbh\ inferred from dynamical modeling to the observed virial product, and $\theta$ is inferred from the same modeling.
For the same sample, $\langle f\rangle$ was constrained by \citet{Wang26}: $\langle \log f \rangle=0.69\pm0.06$ for the rms line dispersion and $\langle \log f \rangle=0.58\pm0.06$ for the mean line dispersion, which are almost equivalent to those constrained by \citet{Woo15}. Thus, adopting this value rather than the value from \citet{Woo15} does not significantly affect our results.

For each AGN in the sample of \citet{Villafana26}, we generate paired Monte Carlo samples of $f$ and $\theta$. We then assign each Monte Carlo draw a Gaussian weight according to the deviation of its $\log f$ from the adopted $\langle \log f\rangle$, such that draws closer to $\langle \log f\rangle$ receive larger weights. The weighted $\theta$ draws from all AGN are then combined into a single distribution, from which we adopt the median $\langle\theta\rangle = 28.8^\circ$ as the representative viewing angle.

\subsection{Modeling $\theta$-Dependence}
\label{sec:m_lambda_calibration}

%We first model the viewing angle dependence of $M_{obs}$ and $\lambda_{obs}$. This step has two roles in the inference. First, for each trial value of $\theta$, it maps $M_{obs}$ and $\lambda_{obs}$ to the corresponding intrinsic parameters, $M$ and $\lambda$, used as FRADO inputs. Second, it defines the expected $\lambda_{obs}$--$\theta$ relation used to construct the conditional viewing angle weights, later to be applied to the likelihood function. Rather than attempting a full physical treatment of the $\theta$-dependence in the FRADO BLR geometry and kinematics, we adopt a phenomenological calibration ansatz for $M_{obs}$ and $\lambda_{obs}$, which is then incorporated into the Bayesian inference of $\theta$. 

First, we model the viewing angle dependence of the time lag. Since the far-side BLR dominates the transfer function (\S~2.2), we assume the lag increases with the viewing angle as $(1+\sin\theta)$ for a thin BLR disk. By accounting for the shape of the transfer function (see \figref{tf}), we generalize this dependence as $(1+\beta\sin\theta)$, where the dimensionless factor $\beta$ captures variations in the transfer function shape arising from the extended BLR geometry in the FRADO model. 
%We parameterize $\beta$ as a function of AGN luminosity.
Then, the dependence of $\tau_{model}$ is expressed as
\begin{equation}
\label{eq:tau_true}
\frac{\tau_{model}(\theta)}{\tau_{model}(\langle \theta \rangle)}
=
\frac{1+\beta \sin\theta}{1+\beta \sin\langle \theta \rangle}.
\end{equation}
%Here, $\langle\theta\rangle$ denotes the representative viewing angle  

Second, we adopt the prescription of \citet{Collin06} for the line velocity by combining rotational and turbulent components. If the rotation is dominant, the projected LOS velocity scales approximately as $\sin\theta$ and the corresponding mass scales as $\sin^2\theta$. In addition, we include a turbulent velocity component related to the BLR thickness (or opening angle). 
Following \citet{Collin06}, we parameterize the ratio of the turbulent velocity to the local Keplerian rotational velocity with the BLR aspect ratio.
%parameterized the contribution of the turbulent velocity component using the ratio of the turbulent velocity to the local Keplerian rotational velocity and identified this ratio with the BLR aspect ratio. 
%Motivated by this prescription, 
We describe the aspect ratio as $\tan\phi$, where $\phi$ is the BLR half-opening angle, and introduce $f_{turb}$ as a free parameter that determines the strength of the turbulent contribution to $\Delta V_{los}^{2}$. Therefore, $\Delta V_{los}^2$ scales as $\sin^2\theta + f_{turb}\tan^2\phi$. Because the calibrated mass is inversely proportional to this factor, a nonzero $f_{turb}$ prevents it from diverging when $\theta$ approaches zero. We constrain $f_{turb}$ through Bayesian inference and treat it as a global parameter common to all AGN in our \Hb\ RM sample. 
We discuss the implications of this parameter further in Section~\ref{sec:openingangle}.
% and defer a more detailed treatment to future work.

Third, combining the dependences of $\tau_{model}$ and $\Delta V_{los}$, we describe the $\theta$-dependence on mass as
\begin{equation}
\label{eq:M_true}
\frac{M}{M_{obs}} =
\frac{\sin^2\langle \theta \rangle + f_{turb} \tan^2\phi}
{\sin^2\theta + f_{turb} \tan^2\phi}\cdot\frac{1+\beta\sin\langle \theta \rangle}
{1+\beta\sin\theta}.
\end{equation}

%The disk luminosity observed at 5100\AA\ is often calculated from the measured flux assuming isotropic emission. However, the disk emission is anisotropic to the LOS. 
Fourth, under a simplified thin Lambertian disk prescription, we assume that the observed continuum luminosity scales with $\cos\theta$, which represents the projected emitting area. Therefore, the calibration of \Lmon\ is given as
\begin{equation}
\label{eq:L5100}
\begin{split}
\frac{L_{5100}}{L_{5100,\ obs}} =\frac{\cos\langle \theta \rangle}{\cos\theta}.
\end{split}
\end{equation}

Finally, by combining Equations~\ref{eq:M_true} and \ref{eq:L5100}, we obtain the Eddington ratio prescription as
\begin{equation}
\label{eq:lambda_true}
\frac{\lambda}{\lambda_{obs}} =
\frac{cos\langle \theta \rangle}{cos\theta}\cdot
\frac{\sin^2\theta + f_{turb} \tan^2\phi}
{\sin^2\langle \theta \rangle + f_{turb} \tan^2\phi}\cdot\frac{1+\beta\sin\theta}
{1+\beta\sin\langle \theta \rangle}
\end{equation}
\\
%Note that Equations~\ref{eq:tau_true}--\ref{eq:lambda_true} are normalized at $\theta=\langle\theta\rangle$, such that $\tau_{model}$, $M$, $L_{5100}$, and $\lambda$ remain unchanged at $\theta=\langle\theta\rangle$.
We parameterize the opening angle $\phi$ and shape factor $\beta$ as functions of the intrinsic AGN luminosity derived from the FRADO model as described in Section~\ref{sec:infmodel}. Note that $\phi$ is defined as the altitude range that includes 95$\%$ of the H$\beta$ emission. 
%Because $\phi$ is tightly correlated with the intrinsic AGN luminosity, we parameterize it as a function of $log \ (M\lambda)$. On the other hand, we determine $\beta$ by fitting the viewing angle dependence of $\tau_{model}$.
%We likewise parameterize $\beta$ as a function of $log \ (M\lambda)$. The adopted functional forms are given in Equation~\ref{eq:openingangle} and Equation~\ref{eq:beta}.

\subsection{Conditional $\theta$ Weights for Marginalization}
\label{sec:infmodel}

We construct a conditional weight used to marginalize the likelihood over $\theta$ using Equation~\ref{eq:lambda_true}, which relates $\lambda_{obs}$ to $\lambda$ through the $\theta$ calibration. While $\theta$ is not expected to correlate with $M$ or $\lambda$, the observed values, \Mobs\ and \Eobs\ may depend on $\theta$ as described in the previous section (see Equations~\ref{eq:M_true} and \ref{eq:lambda_true}).
%The corresponding relation between $\lambda_{obs}$, $\lambda$, and $\theta$ is
%\begin{equation}
%\label{eq:lambda_obs}
%\lambda_{obs} =
%\lambda\frac{\cos\theta}{\cos\langle \theta \rangle}\cdot\frac{\sin^2\langle \theta \rangle + f_{turb} \tan^2\phi}
%{\sin^2\theta + f_{turb} \tan^2\phi}\cdot\frac{1+\beta\sin\langle \theta \rangle}
%{1+\beta\sin\theta}
%\end{equation}
%Here, $\lambda$ is an object-specific latent quantity whose value depends on the trial $\theta$.
%{\color{red} same as Eq. 7}

%The intrinsic Eddington ratios of the sample are described by a population distribution of $\lambda$. Through the viewing angle calibration in Equation~\ref{eq:lambda_true}, this 1D population distribution determines the joint distribution between $\lambda_{obs}$ and $\theta$. 

We build an intrinsic distribution of $\lambda$
as an asymmetric split-normal distribution centered at $\log\lambda_c$, with left and right scatters $\sigma_L$ and $\sigma_R$, respectively. 
For a fixed $\lambda$, Equation~\ref{eq:lambda_true} shows that the viewing angle calibration naturally produces an anti-correlation between $\lambda_{obs}$ and $\theta$, because the ratio $\lambda/\lambda_{obs}$ increases monotonically with $\theta$ for the positive values of $f_{turb}$ and $\beta$ relevant to our model. Changing $\lambda$ shifts this relation horizontally in the $\lambda_{obs}$--$\theta$ plane. The parameters $\lambda_c$, $\sigma_L$, and $\sigma_R$ therefore define the family of $\lambda_{obs}$--$\theta$ relations allowed for the population where the central locus is set by $\lambda_{c}$.

The resulting parameter vector $\boldsymbol{\Phi}$ is
\begin{equation}
\boldsymbol{\Phi} = (\lambda_c, \sigma_{L}, \sigma_{R}, f_{turb}).
\end{equation}
We assign uniform priors over the ranges listed in \autoref{table:priors}.
\begin{table}[t]
\centering
\begin{tabular}{ll}
\hline
Parameter & Prior Range \\
\hline
$\lambda_c$ & $(0.001,\,0.1)$ \\
$\sigma_{L}$ & $(0,\,5)$ \\
$\sigma_{R}$ & $(0,\,5)$ \\
$f_{turb}$ & $(0,\,1)$ \\
\hline
\end{tabular}
\caption{Uniform prior ranges adopted for the global parameters in the Bayesian inference.}
\label{table:priors}
\end{table}

For each sampled $\boldsymbol{\Phi}$, we assign each AGN an object-specific conditional weight over the allowed $\theta$ grid. The likelihood is then marginalized over this grid. For the $i$th AGN with an observed Eddington ratio $\lambda_{obs,i}$, we define the intrinsic Eddington ratio $\lambda_{ij}$ at the $j$th viewing angle grid point using Equation~\ref{eq:lambda_true}:
\begin{equation}
\lambda_{ij} \equiv
\lambda_{\mathrm{obs},i}
\frac{g_{ij}}{\langle g_{ij} \rangle}\cdot
\frac{\cos \langle \theta \rangle }{\cos\theta_j}\cdot
\frac{1+\beta_{ij}\sin\theta_j}
    {1+\beta_{ij}\sin \langle \theta \rangle },
\end{equation}
where
\begin{equation}
\begin{aligned}
g_{ij} &\equiv \sin^2\theta_j + f_{turb}\tan^2\phi_{ij}, \\
\langle g_{ij} \rangle &\equiv \sin^2\langle \theta \rangle
+ f_{turb}\tan^2\phi_{ij}.
\end{aligned}
\end{equation}
Here, $g_{ij}$ represents the combined rotational and turbulent contributions to $\Delta V_{los}^2$ at $\theta_j$, while $\langle g_{ij}\rangle$ denotes the corresponding value at $\langle\theta\rangle$. Moreover, using the FRADO cloud distributions, we calculate $\beta$ and $\phi$ over the full $(M,\lambda)$ grid and obtain linear fits as functions of the luminosity proxy $l_{ij}=M_{ij}\lambda_{ij}$:
\begin{equation}
\label{eq:openingangle}
\phi_{ij} \ (deg)
=
6.85\,\log l_{ij}
-31.4,
\end{equation}
\begin{equation}
\label{eq:beta}
\beta_{ij}
=
0.196\,\log l_{ij}
+0.115.
\end{equation}

For each $\lambda_{ij}$ evaluated at $\theta_j$, we assign the $i$th AGN a (unnormalized) conditional weight $w_{ij}$. This is defined as
\begin{equation}
{w}_{ij} \propto
\sin\theta_j
\left\{
\begin{aligned}
&\exp\!\left[
-\frac{1}{2}
\left(
\frac{\log \lambda_{ij}-\log \lambda_c}{\sigma_{L}}
\right)^2
\right]
\left|\dfrac{d\log \lambda}{d\theta}\right|_{\theta=\theta_j},
\\
&\qquad \text{if } \log \lambda_{ij} < \log \lambda_c,
\\[8pt]
&\exp\!\left[
-\frac{1}{2}
\left(
\frac{\log \lambda_{ij}-\log \lambda_c}{\sigma_{R}}
\right)^2
\right]
\left|\dfrac{d\log \lambda}{d\theta}\right|_{\theta=\theta_j},
\\
&\qquad \text{if } \log \lambda_{ij} \ge \log \lambda_c.
\end{aligned}
\right.
\end{equation}
The factor $\sin\theta_j$ represents the isotropic viewing angle prior expected for randomly oriented systems in three-dimensional space (\citealt{Lawrence10}). Meanwhile, the Jacobian term $\left|d\log\lambda/d\theta\right|$ converts the population weight defined in $\log\lambda$ space into the corresponding weight in $\theta$ space. We normalize the weights as
\begin{equation}
\tilde{w}_{ij} =
\frac{w_{ij}}{\sum_k w_{ik}\Delta_\theta},
\end{equation}
so that
\begin{equation}
\sum_j \tilde{w}_{ij}\Delta_\theta=1.
\end{equation}
Here, $\Delta\theta=0.5^\circ$ denotes the grid spacing in $\theta$. For consistency with the FRADO model grid, we restrict the viewing angle to $[0^\circ ,60^\circ]$.

\subsection{Marginalized Likelihood and Posterior}
\label{sec:likelihood}

Using the conditional viewing angle weights defined in Section~\ref{sec:infmodel}, we construct the $\theta$-marginalized likelihood of \Tobs\ and combine it with the priors on $\boldsymbol{\Phi}$ to obtain the posterior. For each AGN and viewing angle grid point $\theta_j$, we apply the calibration described in Section~\ref{sec:m_lambda_calibration} to obtain $M$ and $\lambda$ from \Mobs\ and \Eobs. These quantities, together with $\theta_j$, are then used to evaluate $\tau_{model,i}(\theta_j)$ and its uncertainty. For each AGN $i$ and viewing angle grid point $\theta_j$, we define the likelihood by modeling $\log \tau_{obs,i}$ as a Gaussian variable centered on $\log \tau_{model,i}(\theta_j)$,
\begin{equation}
\log  \tau_{{ obs},i}\,\big|\,\theta_j,\boldsymbol{\Phi}
\sim
\mathcal{N}\!\left(
\log  \tau_{{ model},i}(\theta_j),\ 
\sigma_{ij}^{2}
\right).
\end{equation}
where
\begin{equation}
\sigma_{ij}^2 \equiv 
\sigma_{\tau,{obs},i}^{2}
+
\sigma_{\tau,{model},i}^{2}(\theta_j).
\end{equation}
Here, $\sigma_{\tau,obs,i}$ and $\sigma_{\tau,model,i}(\theta_j)$ denote the uncertainties of $\tau_{obs}$ and $\tau_{model}$ in log-space, respectively. The likelihood is then marginalized over the $\theta$ grid,
\begin{equation}
p\!\left(\log \tau_{\mathrm{obs},i}\mid\boldsymbol{\Phi}\right)
\approx
\sum_j \tilde{w}_{ij}\,
p\!\left(\log \tau_{\mathrm{obs},i}\mid\theta_j,\boldsymbol{\Phi}\right)
\Delta_\theta
\end{equation}
where $\widetilde{w}_{ij}$ is the normalized conditional weight for the $i$th AGN at $\theta_j$. The conditional likelihood is given by
\begin{equation}
p\!\left(\log  \tau_{{ obs},i}\mid \theta_j,\boldsymbol{\Phi}\right)
=
\frac{1}{\sqrt{2\pi\,\sigma_{ij}^{2}}}
\exp\!\left(-\frac{r_{ij}^2}{2\sigma_{ij}^2}\right),
\end{equation}
where
\begin{equation}
\begin{aligned}
r_{ij} &\equiv 
\log  \tau_{{ obs},i}
-
\log  \tau_{{ model},i}(\theta_j). \\
\end{aligned}
\end{equation}

The total log-likelihood of the sample is therefore
\begin{equation}
\log  \mathcal{L}(\boldsymbol{\Phi})
=
\sum_{i}
\log  
p\!\left(\log  \tau_{{ obs},i}\mid\boldsymbol{\Phi}\right).
\end{equation}

Combining the likelihood with the prior on $\boldsymbol{\Phi}$ yields the posterior distribution
\begin{equation}
p(\boldsymbol{\Phi}\mid \{\log  \tau_{{ obs},i}\})
\propto
p(\boldsymbol{\Phi})\,\mathcal{L}(\boldsymbol{\Phi}).
\end{equation}
In addition, we account for the measurement uncertainties in $\log M_{obs}$ and $\log\lambda_{obs}$ by drawing 25 Gaussian realizations for each AGN and marginalizing the likelihood over these realizations. For each realization, we recompute the viewing angle calibration, the conditional viewing angle weights, and $\tau_{model}$. The final likelihood is obtained by averaging the $\theta$-marginalized likelihood over the 25 realizations.

We sample the posterior using the affine-invariant ensemble sampler implemented in the \texttt{emcee} package. The resulting posterior samples constrain the four model parameters and are used to infer $\theta$ of each AGN. We then propagate these posterior samples to obtain the corresponding posterior distributions of $M$, $\lambda$, and posterior-predictive model time lag, $\tau_{model}$. 

%%%%%%%%%%%%%%%%%%%%%%%%%%%%%%%

\section{Results}
\label{sec:results}

In this section, we present the main results obtained by applying the Bayesian framework to the \Hb\ RM sample of \citet{Wang24}. First, we describe the posterior constraints on the model parameters and the resulting distributions of $\theta$, $M$, and $\lambda$ for the \Hb\ RM sample (\S~\ref{sec:theta_posterior}). Second, we investigate the sample selection effect on the inferred $\theta$--$\lambda$ and $\theta$--$M$ relations (\S~\ref{sec:mocksample}).
%and the joint distributions of the inferred viewing angles with the calibrated $M$ and $\lambda$. 
Third, we compare the inferred $\tau_{model}$ with the observed H$\beta$ lags and evaluate the reproduced \RBLR--\Lmon\ relation against observations (\S~\ref{sec:modelRLrel}).

%\subsection{Viewing Angles}
\subsection{Posterior Constraints}
\label{sec:theta_posterior}

\begin{figure*}[!t]
  \centering
  \begin{subfigure}[b]{0.48\textwidth}
    \centering
    \includegraphics[height=0.93\textwidth]{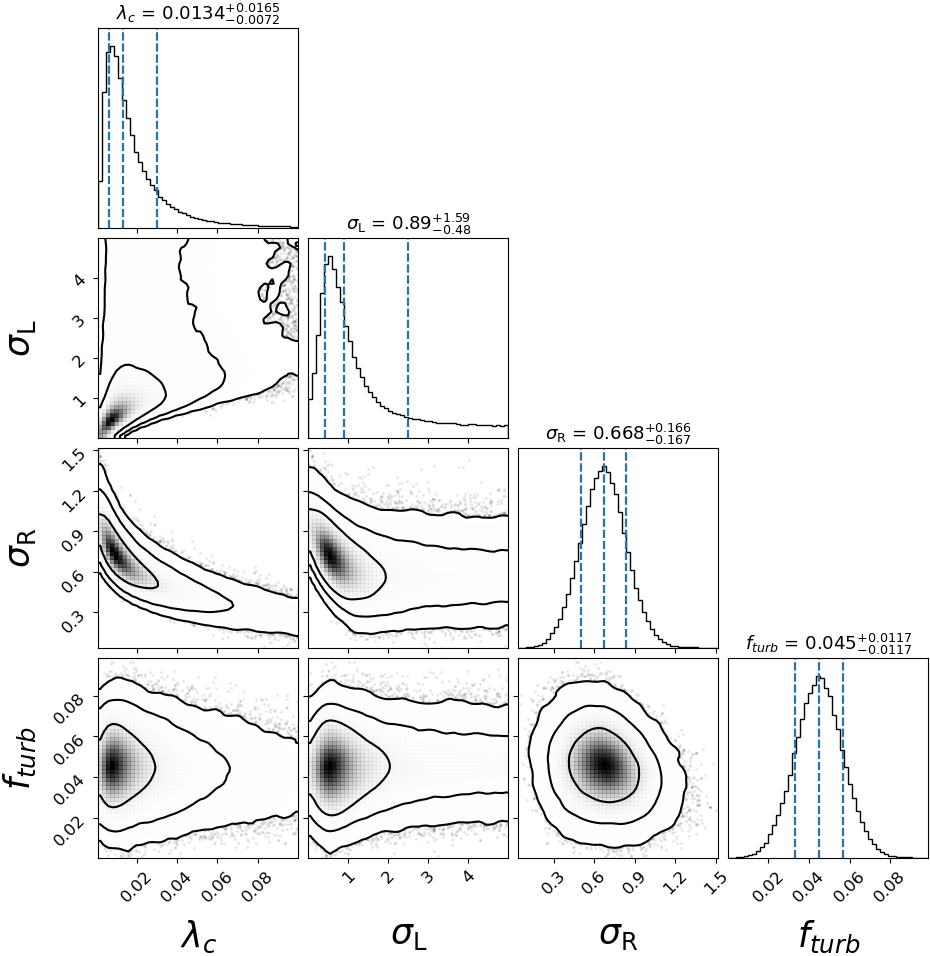}
    \phantomsubcaption
    \label{posterior(a)}
  \end{subfigure}\hfill%
  \begin{subfigure}[b]{0.48\textwidth}
    \centering
    \includegraphics[height=0.93\textwidth]{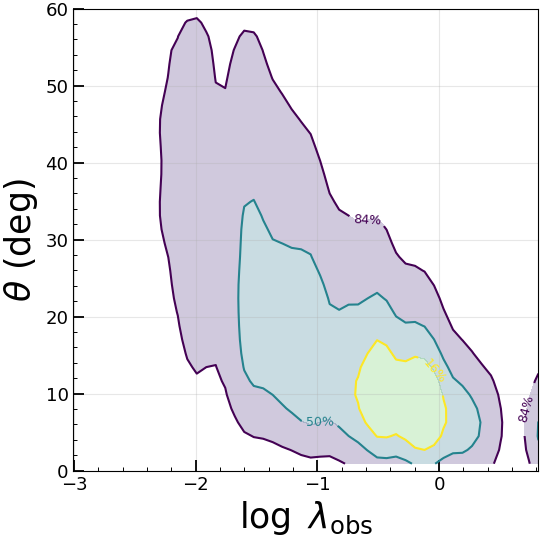}
    \phantomsubcaption
    \label{posterior(b)}
  \end{subfigure}
  \caption{Left: Joint posterior distributions of the four model parameters. Regions enclosing 68$\%$, 95$\%$, and 99.7$\%$ are denoted as contours, while the marginalized 1D posterior distributions are shown in the diagonal panels. The 16th, 50th, and 84th percentiles of the 1D distributions are indicated by the blue dashed vertical lines. Right: 2D posterior probability distribution of the observed Eddington ratio ($\lambda_{obs}$) and the inferred viewing angle ($\theta$), showing the conditional distribution of $\theta$ at a given $\lambda_{obs}$. Regions enclosing 16$\%$, 50$\%$, and 84$\%$ are denoted as contours.}
  \label{posterior}
\end{figure*}

Applying the Bayesian framework to the \Hb\ RM sample, we obtain the posterior constraints on the model parameters and the viewing angles as shown in  \figref{posterior(a)}. 
From the joint posterior distributions of the four parameters, we obtain the 1D parameter constraints as
$\lambda_c = 0.013^{+0.017}_{-0.0072}$,
$\sigma_{L} = 0.89^{+1.59}_{-0.48}$,
$\sigma_{R} = 0.67^{+0.17}_{-0.17}$,
and $f_{turb} = 0.046^{+0.012}_{-0.012}$ with 
% 16th--84th percentile ranges. 
1-$\sigma$ uncertainties. 
%In addition to these marginalized constraints, 
Note that these posterior constraints depend on the input sample. Thus, these parameters are relevant for
the \Hb\ RM sample. 
By defining the maximum a posteriori (MAP) parameter vector, $\boldsymbol{\Phi_{MAP}}$, 
%as the parameter vector 
that maximizes the posterior probability in the full four-dimensional parameter space, we obtain
$\lambda_c=0.0063$, 
$\sigma_{L}=0.36$, 
$\sigma_{R}=0.78$,
and $f_{turb}=0.046$.
We adopt $\boldsymbol{\Phi_{MAP}}$ as the representative parameter vector for the fixed-$\boldsymbol{\Phi}$ analyses in Section~\ref{sec:modelRLrel}, which assess the sensitivity of the results to uncertainty in $\boldsymbol{\Phi}$.

Using the posterior constraints on $\boldsymbol{\Phi}$, we infer the $\theta$ distribution for each AGN in the \Hb\ RM sample. We present the resulting family of $\lambda_{obs}$--$\theta$ relations
% implied by the posterior distribution of $\boldsymbol{\Phi}$ 
with a probability density map in \figref{posterior(b)}. 
Despite the broad dispersion in $\theta$ at fixed $\lambda_{obs}$, AGN with higher $\lambda_{obs}$ preferentially 
have smaller viewing angles.

\begin{figure*}[!t]
  \centering
  \begin{subfigure}[t]{0.46\textwidth}
    \vspace{0pt}\centering
    \includegraphics[height=0.94\textwidth]{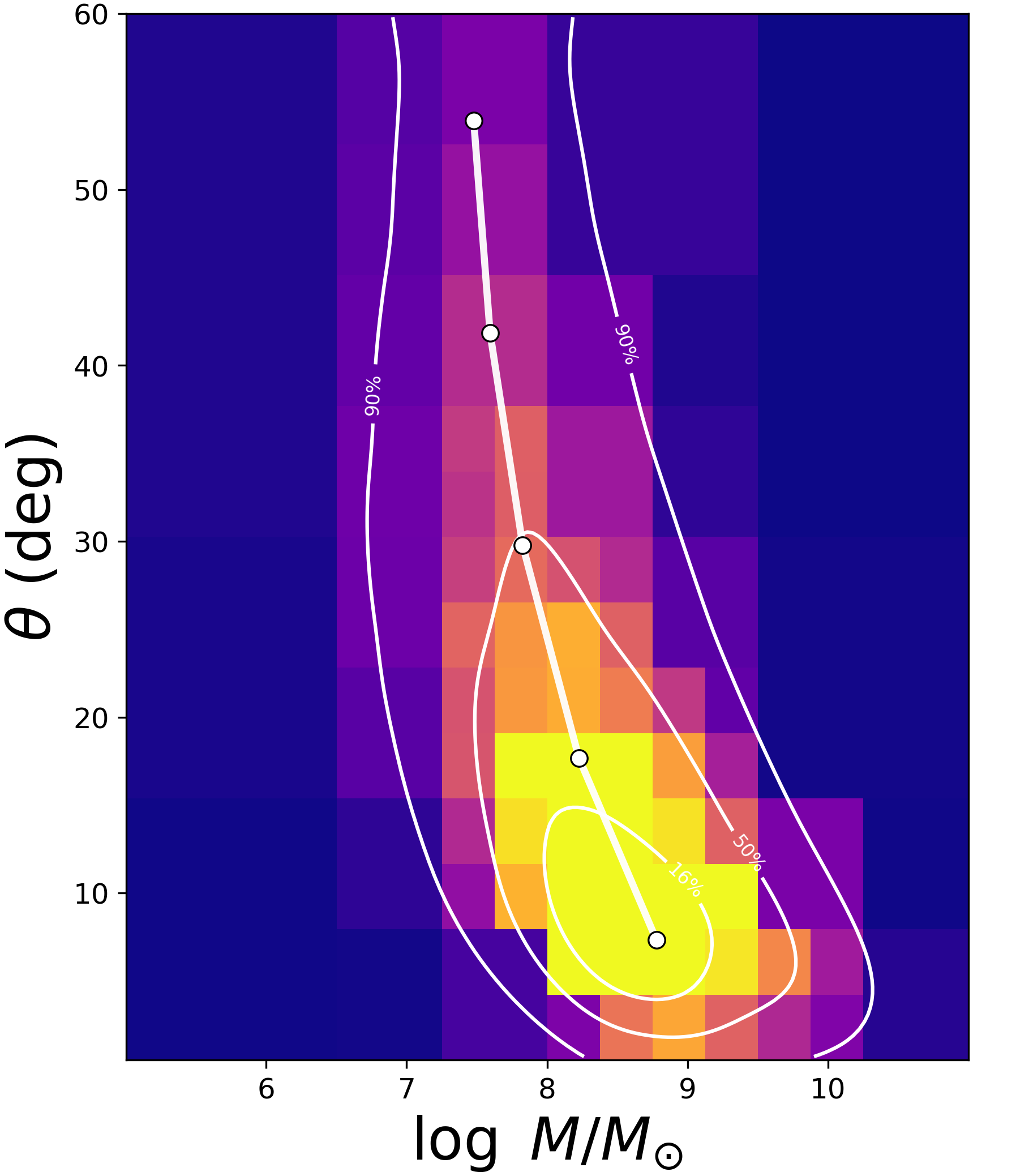}
    \phantomsubcaption
  \label{theta_heatmap(a)}
  \end{subfigure}
  \begin{subfigure}[t]{0.46\textwidth}
    \vspace{0pt}\centering
    \includegraphics[height=0.94\textwidth]{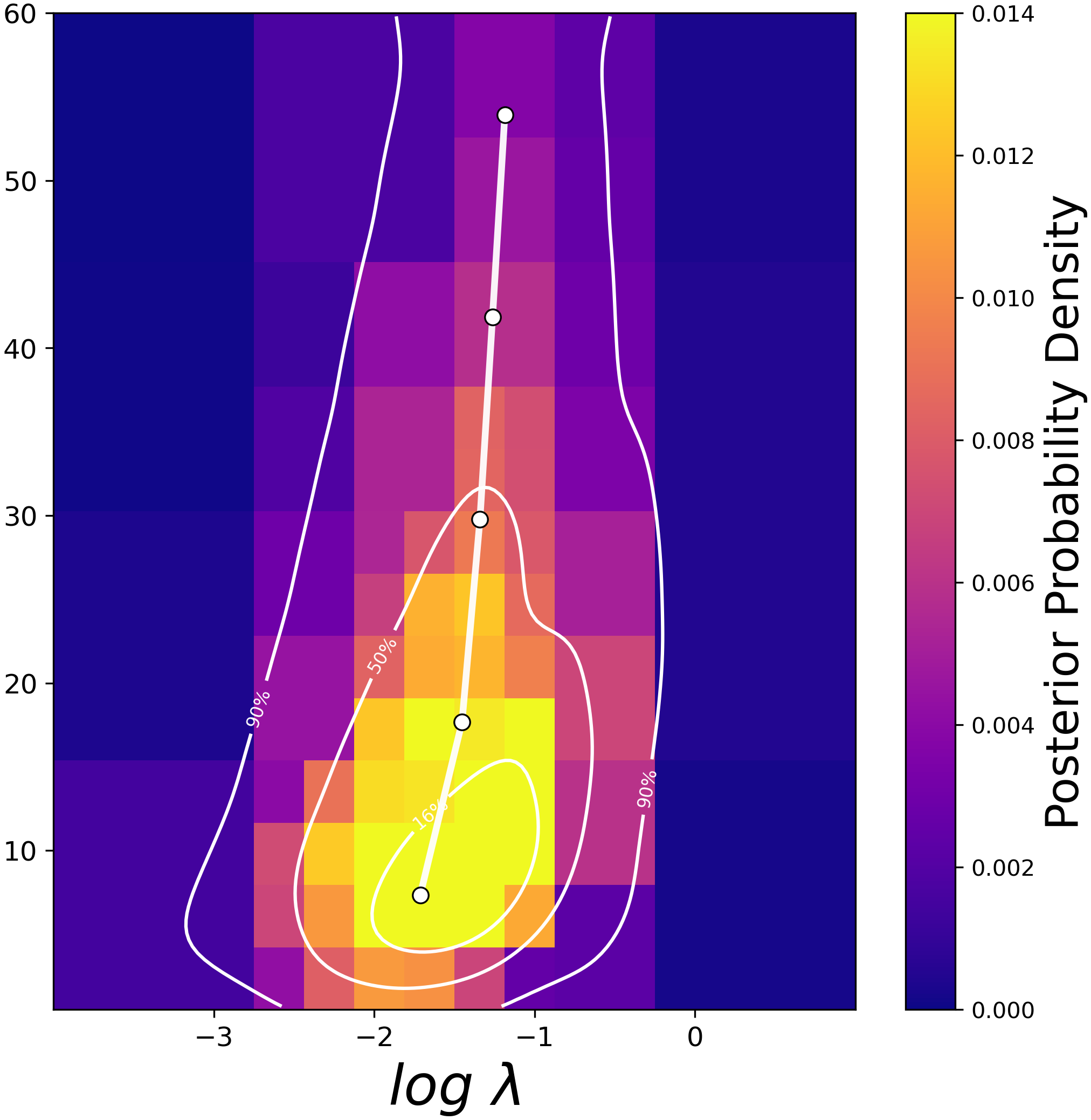}
    \phantomsubcaption
  \label{theta_heatmap(b)}
  \end{subfigure}
  \caption{Joint posterior distributions of the viewing angle and the intrinsic \mbh\ (left) and Eddington ratio (right). Regions enclosing 16$\%$, 50$\%$, and 90$\%$ are denoted as contours. The centroid trends are indicated by the white lines.}
  \label{theta_heatmap}
\end{figure*}

Then, we construct the joint posterior distributions of the inferred $\theta$ and the calibrated $M$ and $\lambda$ 
by drawing 200 samples of $\boldsymbol{\Phi}$ from the posterior and generating 200 $\theta$ realizations for each sampled vector. For each $\theta$ realization, we obtain the corresponding $M$ and $\lambda$ through the viewing angle calibration (see \S~\ref{sec:m_lambda_calibration}).
%The resulting joint distributions are visualized as heatmaps in \figref{theta_heatmap}.
We present the resulting distributions in \figref{theta_heatmap}.
Note that $M$ and $\lambda$ are expected to be independent of $\theta$. 
%no systematic trend with viewing angle, 
However, the posterior contours exhibit residual trends with $\theta$, which we attribute to the incompleteness of the \Hb\ RM sample
% rather than an intrinsic physical relation, 
(see \S~\ref{sec:mocksample}).

\begin{figure}[!t]
  \centering
    \vspace{0pt}
    \centering
    \includegraphics[width=0.93\linewidth]{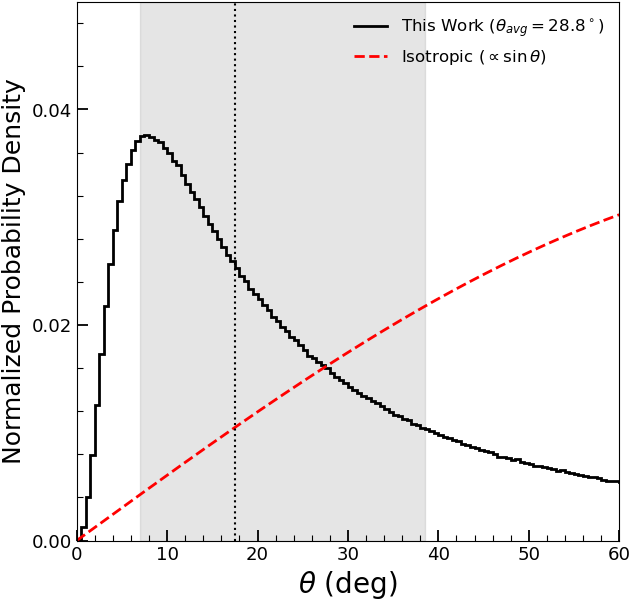}
    \caption{Marginalized posterior probability distribution of $\theta$ (black line), constructed using the sampling configuration of \figref{theta_heatmap}. 
    %The distribution inferred in this work is shown by the black solid line. 
    The 16th--84th percentile range is indicated by the shaded gray region, while the median is marked by the vertical dotted line. For comparison, the $\theta$ distribution expected for an isotropic orientation in three-dimensional space, $p(\theta)\propto\sin\theta$, is shown by the red line.}
  \label{theta_pdf}
\end{figure}

For the \Hb\ RM sample, we obtain a marginalized $\theta$ distribution with a mean of $21.6^\circ$, which is smaller than the representative viewing angle $\langle\theta\rangle$ corresponding to $\langle f\rangle$ (see \figref{theta_pdf}).
It is also substantially smaller than the isotropic viewing angle prior adopted in our inference, $p(\theta)\propto\sin\theta$.
As the BLR is visible at relatively low viewing angles, this result is qualitatively consistent with the type 1 AGN unification scheme. Nevertheless, the inferred viewing angles are substantially smaller than typically expected for type 1 AGN. We discuss this result further in Section~\ref{sec:theta_discussion}.

\subsection{Selection Effects in the $\theta$--$\lambda$ and $\theta$–$M$ Relations}
%\subsection{Selection Effects in the $\theta$--$\lambda$ and $\theta$--$M$ Relations}
\label{sec:mocksample}

\begin{figure*}[!t]
  \centering
  \begin{subfigure}[t]{0.32\textwidth}
    \vspace{0pt}\centering
    \includegraphics[height=5.9cm,keepaspectratio]{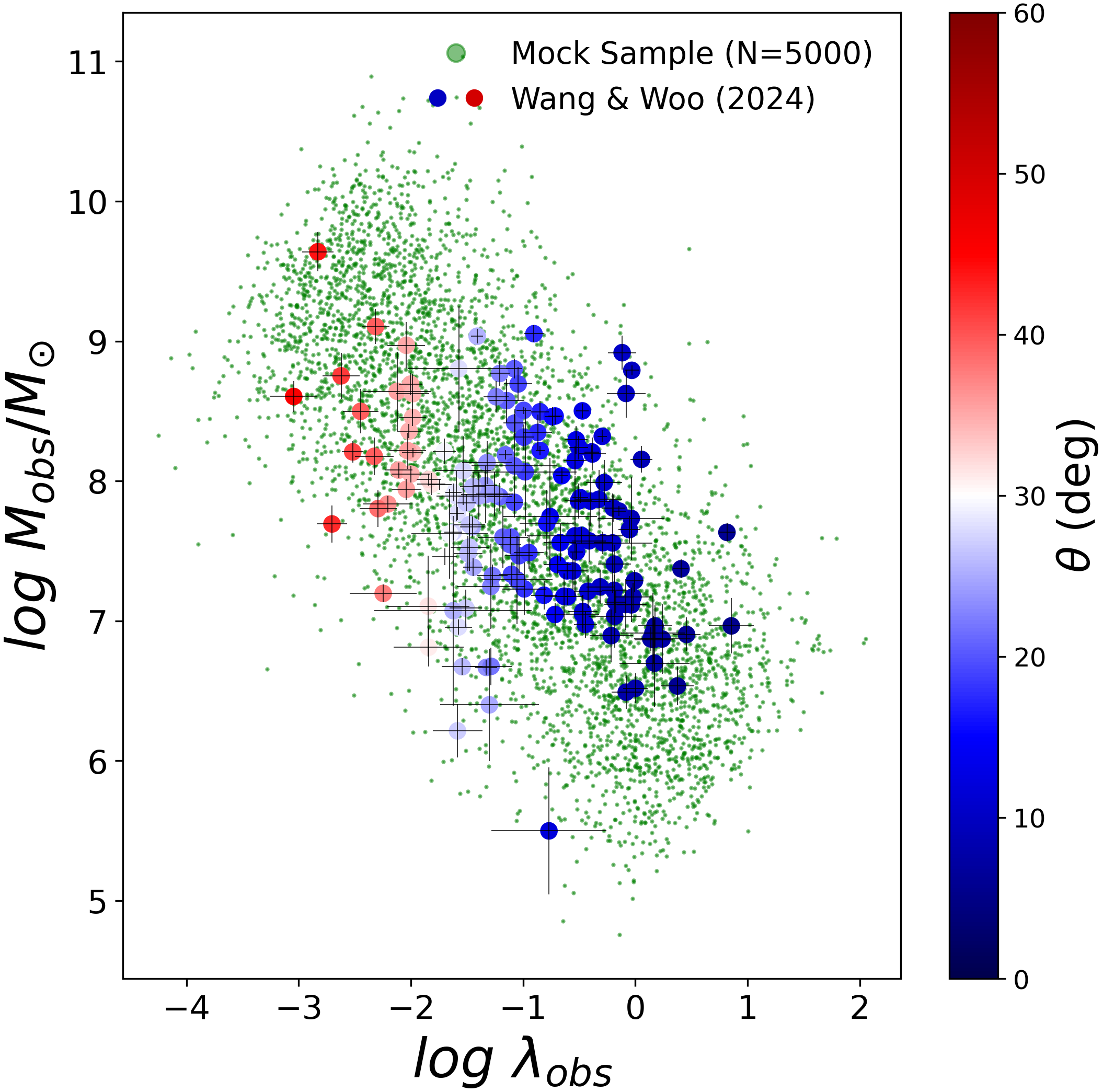}
    \phantomsubcaption
    \label{theta_heatmap_mock(a)}
  \end{subfigure}
  \begin{subfigure}[t]{0.32\textwidth}
    \vspace{0pt}\centering
    \includegraphics[height=5.9cm,keepaspectratio]{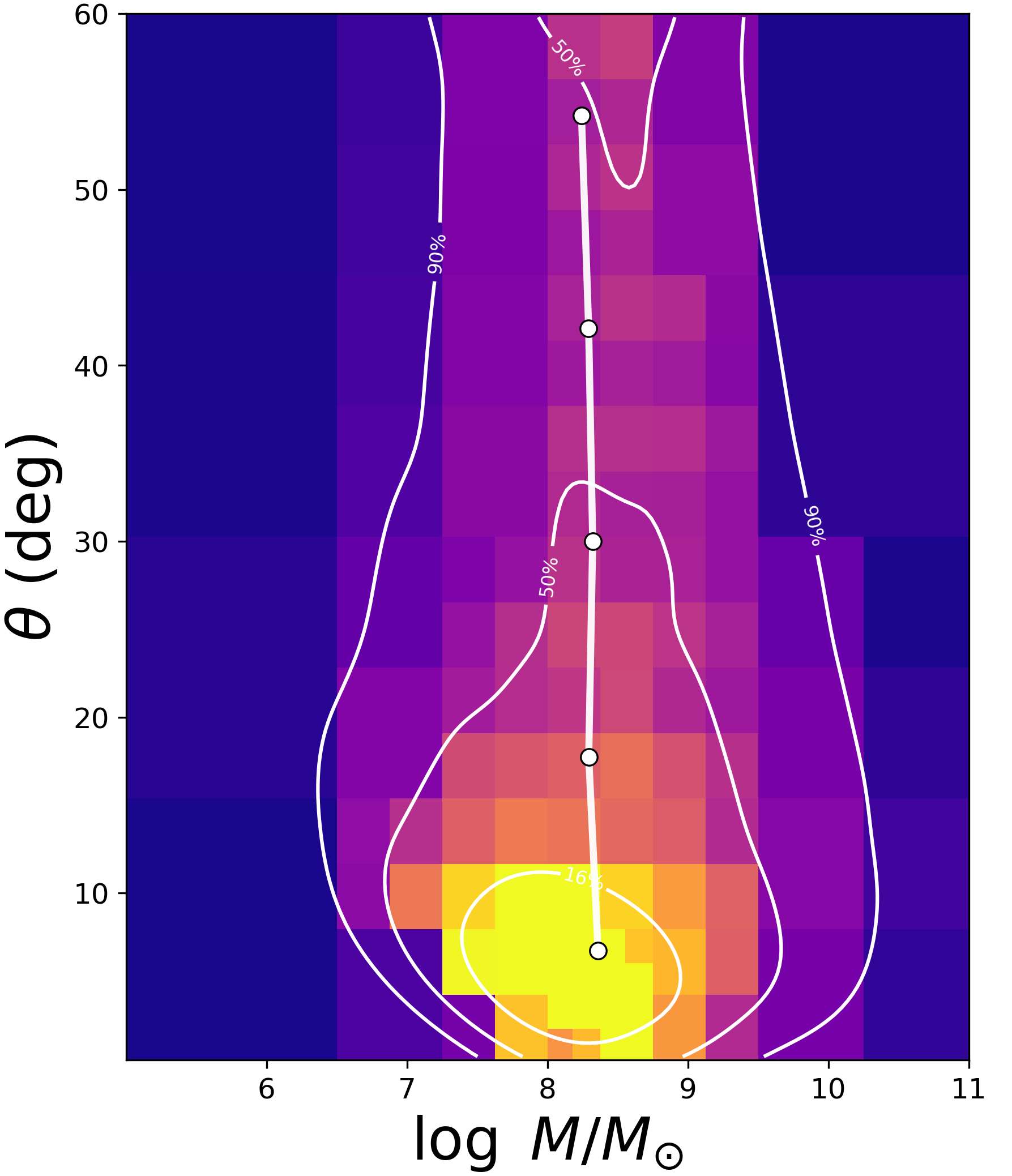}
    \phantomsubcaption
    \label{theta_heatmap_mock(b)}
  \end{subfigure}
  \begin{subfigure}[t]{0.32\textwidth}
    \vspace{0pt}\centering
    \includegraphics[height=5.9cm,keepaspectratio]{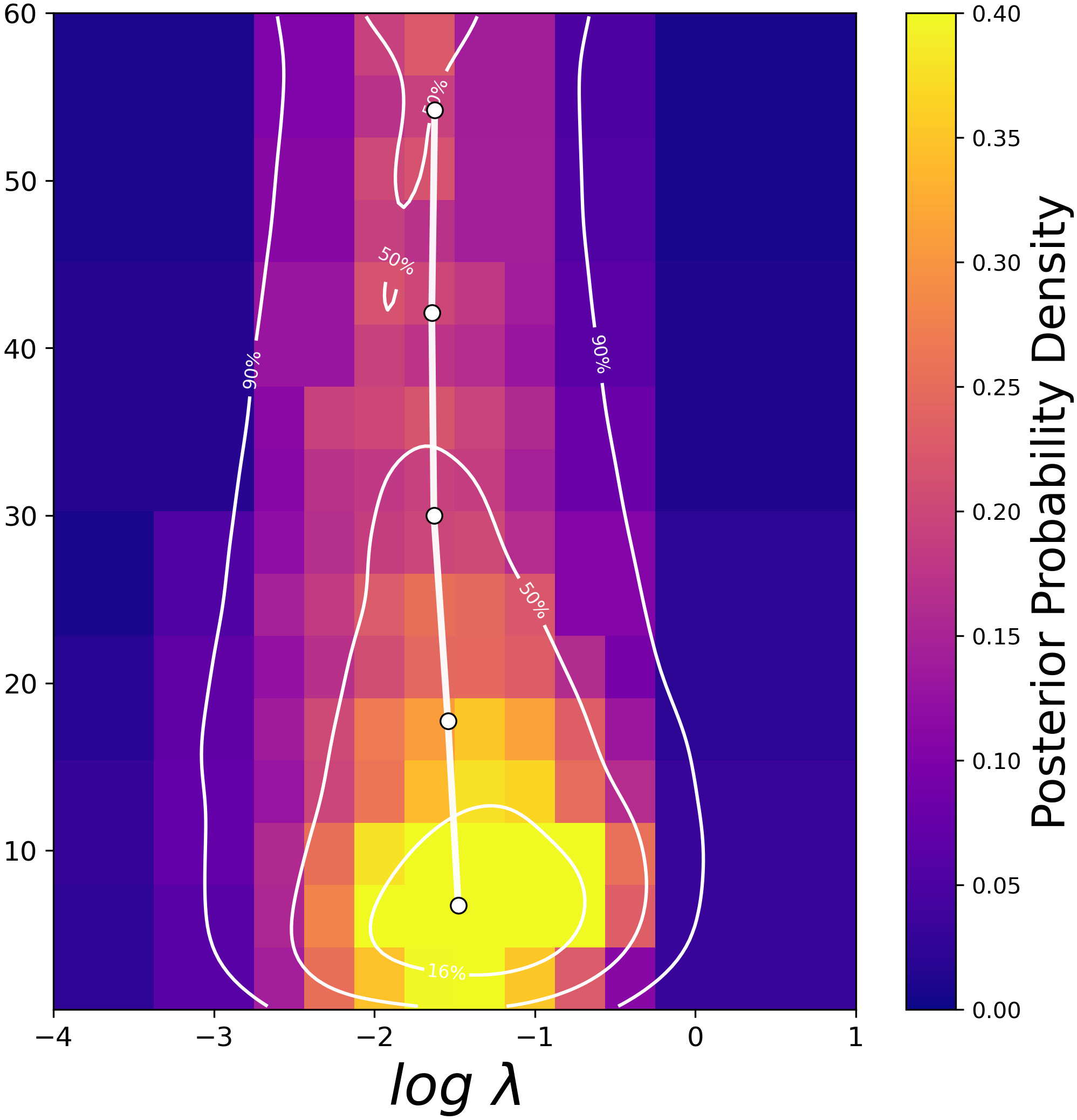}
    \phantomsubcaption
    \label{theta_heatmap_mock(c)}
  \end{subfigure}
\caption{Left: Joint distribution of $\lambda_{obs}$ and $M_{obs}$ for the \Hb\ RM sample, color-coded by the median inferred $\theta$. A mock population of 5000 AGN is shown as green points for comparison. Middle: Joint posterior distribution of $M$ and the inferred $\theta$. Right: Joint posterior distribution of $\lambda$ and the inferred $\theta$. In the middle and right panels, the 16$\%$, 50$\%$, and 90$\%$ posterior-probability regions are shown as contours, and the centroid trends are indicated by the overplotted lines.}
  \label{theta_heatmap_mock}
\end{figure*}

%By propagating the posterior samples of $\theta$,
%through the viewing angle calibrations in Equations~\ref{eq:M_true} and \ref{eq:lambda_true}, 
%we obtain the posterior distributions of $M$ and $\lambda$ for each AGN fwith $M_{obs}$ and $\lambda_{obs}$. 
%Neither $M$ nor $\lambda$ is expected to systematically depend on the viewing angle. However, as shown in \figref{theta_heatmap}, $\lambda$ shows 
Since $\theta$ is expected to be independent of the intrinsic AGN properties, $M$ and $\lambda$, the weak correlations shown in \figref{theta_heatmap} require further investigation. We therefore test whether these trends arise from the limited dynamic range of the \Hb\ RM sample.
To test this interpretation, we generate a mock population of $\sim5000$ AGN such that 
%$\log(\lambda_{\rm obs}\cdot M_{obs})$ 
$\log L_{5100,obs}$
follows the same Gaussian distribution 
%with the mean and scatter measured for 
as the \Hb\ RM sample, 
while $\log\lambda_{obs}$ is uniformly distributed.
% over its observed range.
%Using the mock AGN sample, we revisit the $\theta$--$\lambda$ and $\theta$--$M$ distributions as shown in \figref{theta_heatmap_mock}.

We compare the $M_{obs}$ and $\lambda_{obs}$ distributions of the mock AGN sample and the \Hb\ RM sample in \figref{theta_heatmap_mock(a)}.
The \Hb\ RM sample lacks low-Eddington-ratio AGN at high luminosities and high-Eddington-ratio AGN at low luminosities, whereas the mock sample provides denser coverage of both regimes.
With this more complete coverage in the $\lambda_{obs}$ distribution, 
we now confirm no dependence of $\theta$ on $M$ or $\lambda$, 
%the centroid trends become approximately vertical in both the $\theta$--$\lambda$ and $\theta$--$M$ planes, 
as shown in \figref{theta_heatmap_mock(b)} and \figref{theta_heatmap_mock(c)}.  The residual correlations in \figref{theta_heatmap} are likely caused by the incomplete coverage of the \Hb\ RM sample in the $(\lambda_{obs}, M_{obs})$ plane. Note that the mock sample also shows relatively small $\theta$  since we use the posterior samples of $\boldsymbol{\Phi}$ obtained from the \Hb\ RM sample.
%which also determine the inferred $\theta$ distribution presented in Section~\ref{sec:theta_posterior} and discussed further in Section~\ref{sec:theta_discussion}.

\subsection{\RBLR--\Lmon\ Relation}
\label{sec:modelRLrel}

\begin{figure*}[!t]
  \centering
  \begin{subfigure}[t]{0.48\textwidth}
    \vspace{0pt}\centering
    \includegraphics[height=0.94\textwidth]{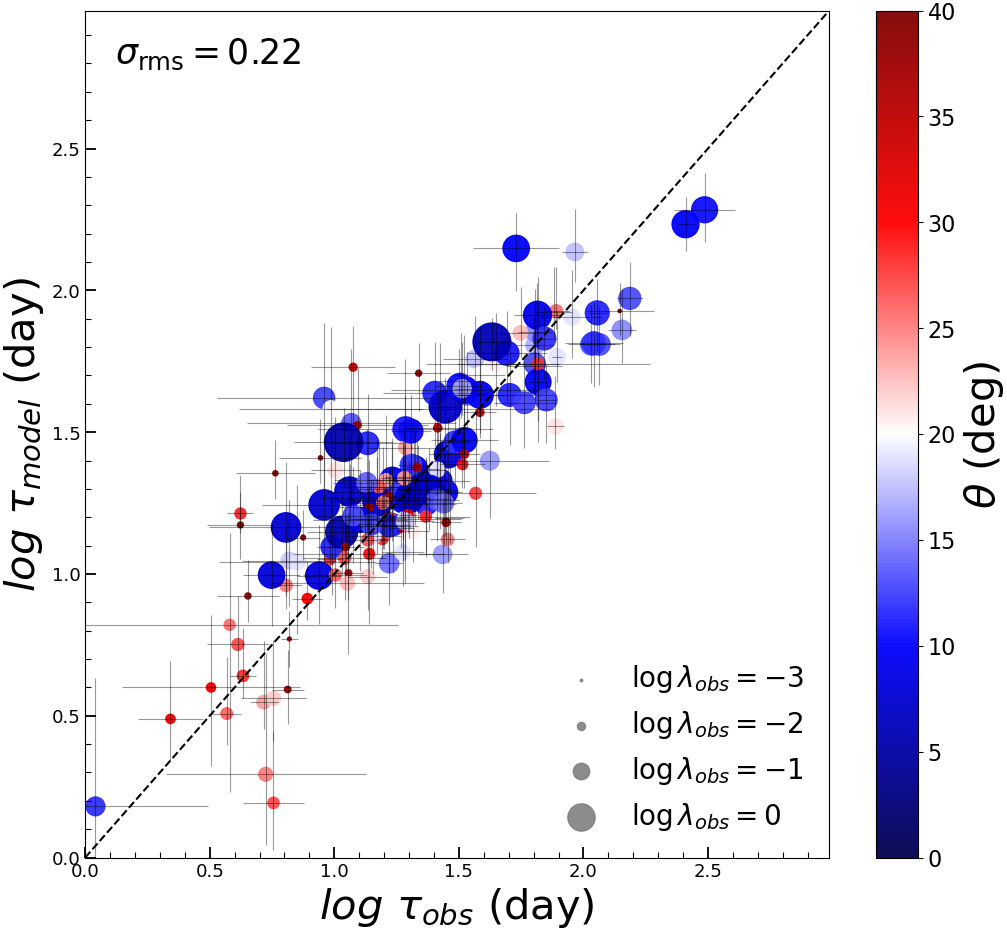}
    \phantomcaption
    \label{RL_taucomp_a}
  \end{subfigure}
  \begin{subfigure}[t]{0.48\textwidth}
    \vspace{0pt}
    \centering
    \includegraphics[height=0.94\textwidth]{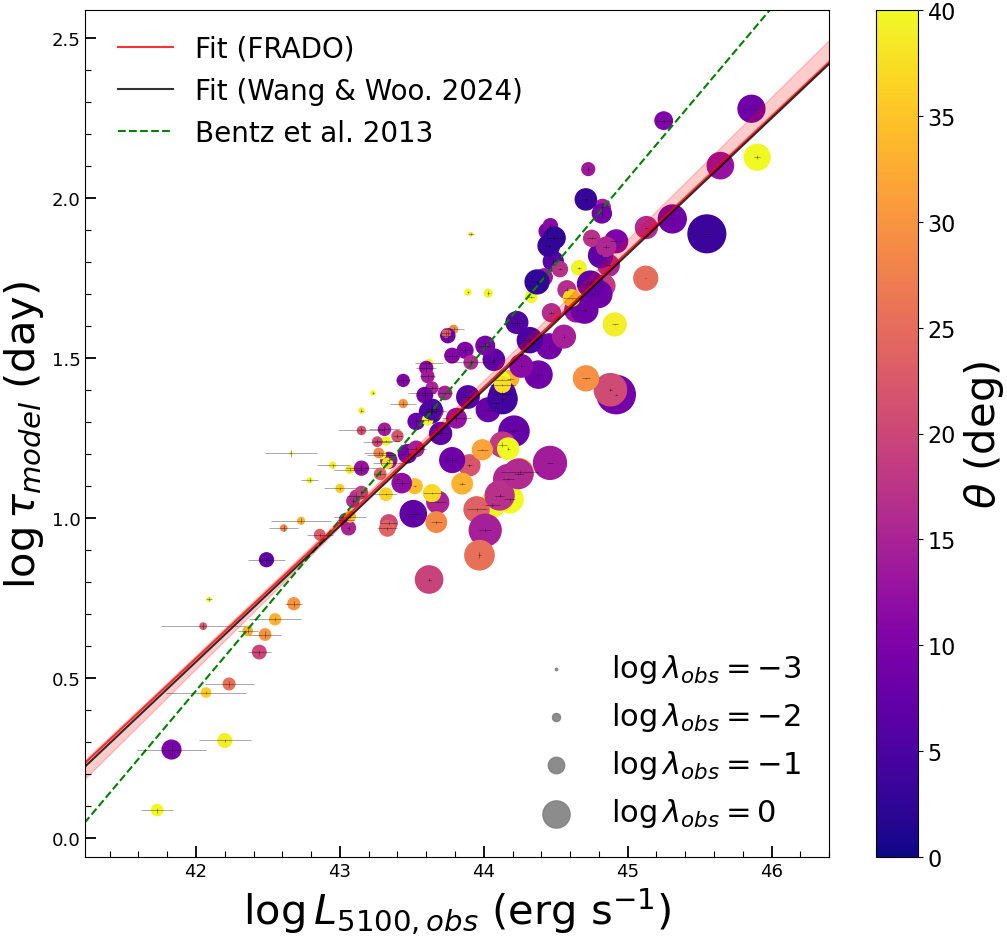}
    \phantomcaption
    \label{RL_taucomp_b}
  \end{subfigure}
  \caption{Left: Comparison of $\tau_{model}$ with $\tau_{obs}$. The error bar of  $\tau_{model}$ indicates the 68$\%$ credible intervals of the posterior-predictive $\tau_{model}$ distributions. The median inferred $\theta$ of each AGN is color-coded. 
 Right: The reproduced `observed' \RBLR--\Lmon\ relation at fixed $\boldsymbol{\Phi_{MAP}}$ with the best-fit (red line) and the 
 1-$\sigma$ uncertainty range obtained from 1000 $\theta$ draws (red shaded region).
  %A single realization based on one $\theta$ draw is shown by the 
The values of $\theta$ and $\lambda_{obs}$ are indicated by color and point size, respectively.  
The best-fit relation for the \Hb\ RM AGN from \citet{Wang24} (black line) and the \citet{Bentz13} relation (green dash-dotted line)
are also presented for comparison.
}
  \label{RL_taucomp}
\end{figure*}

%In the previous section, we inferred the viewing angles of the \Hb\ RM sample by matching $\tau_{model}$ to $\tau_{obs}$. {\color{blue}{In this section, we quantify how well}} $\tau_{model}$ reproduces $\tau_{obs}$ in two complementary ways: an object-by-object comparison of the time lags and a population-level evaluation of the \RBLR--\Lmon\ relation.

Using the sampling configuration described in Section~\ref{sec:theta_posterior}, we draw Monte Carlo samples of $\theta$ for each AGN in the \Hb\ RM sample. 
We apply the viewing angle calibration to each $\theta$ sample to obtain the corresponding $M$ and $\lambda$, and then use $(M,\lambda,\theta)$ to calculate $\tau_{model}$ with the FRADO model (see \S~\ref{sec:lagModeling}). The resulting samples define the posterior-predictive distribution of $\tau_{model}$ for each AGN.

%We then quantify the agreement between $\tau_{model}$ and $\tau_{obs}$. Because we generate the model \RBLR--\Lmon\ relation using the observed $L_{obs}$ values, any offset from the observed relation directly reflects the difference between $\tau_{model}$ and $\tau_{obs}$.

As a consistency check, we compare $\tau_{model}$ with $\tau_{obs}$ (\figref{RL_taucomp_a}), finding consistency between them with a negligible 
offset of 0.037 dex and an rms scatter of 0.22dex, indicating that we properly reproduced the observed time lag.
%The intrinsic scatter from the one-to-one relation is $\sim0.079$ dex.
%further demonstrating the agreement between $\tau_{model}$ and $\tau_{obs}$ under the inferred viewing angle distribution.
To test the effect of the uncertainty in $\boldsymbol{\Phi}$, we fix the parameter vector at $\boldsymbol{\Phi_{MAP}}$ and generate 1000 viewing angle realizations. We find that the $\tau_{model}$ is consistent with $\tau_{obs}$ with an offset of 0.042 dex and an rms scatter of 0.21 dex, indicating that uncertainty in $\boldsymbol{\Phi}$ has little effect on the result. We also repeat the analysis without the measurement uncertainties in \Mobs\ and $\lambda_{obs}$, finding a comparable result.
% $\tau_{obs}$--$\tau_{model}$ distribution remains similar, although the intrinsic scatter increases slightly to 0.096 dex.

Then, we test whether our model can reproduce the observed \RBLR--\Lmon\ relation. We first fix the parameter vector at $\boldsymbol{\Phi_{MAP}}$ and {generate 1000 Monte Carlo realizations of $\theta$ for the \Hb\ RM sample. For each realization, we measure the slope and intrinsic scatter and construct their probability distributions. 
We obtain an \RBLR--\Lmon\ slope of $0.44^{+0.011}_{-0.010}$ and an intrinsic scatter of $0.19^{+0.0078}_{-0.0085}$ dex using the 1000 realizations, where the uncertainties represent the 1-$\sigma$ ranges, finding that the model properly reproduces the observed \RBLR--\Lmon\ relation reported by \citet{Wang24}. 
%differing by $\sim0.67\sigma$ in slope and $\sim0.61\sigma$ in intrinsic scatter.
We repeat the analysis by drawing $\boldsymbol{\Phi}$ from its full posterior distribution using the same sampling configuration from Section~\ref{sec:theta_posterior}. We obtain an \RBLR--\Lmon\ slope of $0.43^{+0.016}_{-0.017}$ and an intrinsic scatter of $0.19^{+0.016}_{-0.015}$ dex, which are in better agreement with those measured by \citet{Wang24}.
%differing by only $\sim0.25\sigma$ and $\sim0.29\sigma$, respectively.}}

In \figref{RL_taucomp_b}, we illustrate the observed \RBLR--\Lmon\ relation generated from a single $\theta$ draw at fixed $\boldsymbol{\Phi_{MAP}}$. AGN with higher $\lambda$ tend to lie below the mean \RBLR--\Lmon\ relation, qualitatively consistent with the trend with observed Eddington ratio. However, the calibrated $\lambda$ distribution of the \Hb\ RM sample is systematically shifted toward lower values relative to the $\lambda_{obs}$ distribution of {\citet{Wang24}}. We discuss this further in Section~\ref{sec:calibration_effects}.

We successfully reproduce the shallower slope of the observed R-L relation and the residual trend toward smaller $\tau_{obs}$ for AGN with higher $\lambda_{obs}$ \citep{Wang24}. 
Note that this shallower slope is mainly caused by the viewing angle effect of the \Hb\ RM sample, which has a specific Eddington ratio distribution. Although there is a self-correlation between $\tau_{obs}$ and $\lambda_{obs}$ since $\lambda_{obs}\propto L_{5100,obs}/(\tau_{obs}\Delta V_{los}^{2})$, the offset from the intrinsic R-L relation is not expected to directly correlate with the intrinsic $\lambda$ (see Figure 19 of \citealt{Woo24}).

\section{Discussion}
\label{sec:discussion}

In this section, we discuss the inferred viewing angle distribution of the \Hb\ RM sample (\S~\ref{sec:theta_discussion}), the dependence of our results on the adopted representative viewing angle $\langle\theta\rangle$ (\S~\ref{sec:theta_avg_discussion}), the treatment of the BLR opening angle $\phi$ (\S~\ref{sec:openingangle}), and the differences between the observed AGN properties and their calibrated counterparts (\S~\ref{sec:calibration_effects}).

\subsection{Viewing Angles of the \Hb\ RM Sample}
\label{sec:theta_discussion}

Our Bayesian inference applied to the \Hb\ RM sample favors viewing angles smaller than the representative viewing angle $\langle\theta\rangle=28.8^\circ$, which was derived from the virial factors and viewing angles of individual AGN (see \S~\ref{theta_avg}). As shown in \figref{theta_pdf}, the viewing angle distribution of the sample is skewed toward low viewing angles compared to the isotropic expectation. This trend is also evident in comparison with $\theta$ constrained by dynamical modeling. Overall, the inferred viewing angles of the \Hb\ RM sample are lower than those typically expected for type~1 AGN.

Several studies have suggested that narrow-line Seyfert~1 galaxies (NLS1s), which typically show high Eddington ratios, may preferentially be viewed at low viewing angles (\citealt{Murayama01}; \citealt{Nagao01}; \citealt{Bian04}; \citealt{Decarli08}; \citealt{Peterson11}; \citealt{Sbarrato18}). Although the majority of AGN in the \Hb\ RM sample have \Hb\ FWHM values greater than $2000~{\rm km~s^{-1}}$ and therefore are not classified as NLS1s, the proposed face-on interpretation of high-Eddington-ratio NLS1s is qualitatively analogous to the anticorrelation between $\lambda_{obs}$ and $\theta$ (\figref{posterior(b)}).

%The preference for small viewing angles may also partly reflect  observational selection effects in the current \Hb\ RM sample. AGN observed at smaller viewing angles appear brighter because of the projected disk emission and are less likely to be obscured. Furthermore, they also produce transfer functions with narrower and more prominent peaks, potentially enabling more reliable lag measurements. However, this potential selection effect is not explicitly modeled in the present analysis. This interpretation is qualitatively consistent with the CARAMEL mock distribution shown in \figref{theta_pdf}, which is also shifted toward smaller viewing angles relative to the isotropic expectation, $p(\theta)\propto\sin\theta$.

Compared to the viewing angles from dynamical modeling reported by \citet{Villafana26}, the inferred $\theta$ is lower by $8.5^\circ\pm13^\circ$ for the 31 overlapping AGN. The two estimates are broadly consistent within their uncertainties. However, several objects have inferred viewing angles substantially smaller than the corresponding dynamical modeling estimates.

These differences arise because the inference adjusts $\theta$ to bring $\tau_{model}$ into agreement with $\tau_{obs}$. When $\theta<\langle\theta\rangle$, the viewing angle calibration increases $M$ and decreases $\lambda$ relative to $M_{obs}$ and $\lambda_{obs}$, respectively (see Section~\ref{sec:m_lambda_calibration}). Because $\tau_{model}$ depends strongly on $M$ (see \figref{tau_model}), the increase in $M$ produces a longer $\tau_{model}$. Thus, objects whose initial $\tau_{model}$ based on $(M_{obs}, \lambda_{obs}, \langle\theta\rangle)$ is much shorter than $\tau_{obs}$ favor $\theta$ smaller than $\langle\theta\rangle$. Because this behavior depends on the adopted viewing angle calibration, the current calibration model may require further refinement. In particular, the viewing angle dependence of the line velocity should be modeled more realistically using the FRADO cloud distribution. We defer further calibration to future work.

The discrepancy between the inferred $\theta$ and the dynamical modeling estimates may also arise from the degeneracy between $\theta$ and other BLR parameters, particularly the opening angle $\phi$. In our Bayesian framework, a large $\phi$ weakens the viewing angle dependence of the model time lag as well as the line velocity (see \S~\ref{sec:m_lambda_calibration}), thereby broadening the probability distribution of $\theta$. The preference for small viewing angles is presumably due to the relatively small $\phi$ predicted by the FRADO model: for the \Hb\ RM sample, $\phi$ rarely exceeds $30^\circ$, even for the most luminous AGN (see \S~\ref{sec:openingangle}). The degeneracy among the inferred $\theta$, $M$, and $\phi$ has been reported in BLR dynamical modeling \citep[see also][]{Pancoast14,Grier17a}. In dynamical modeling, this degeneracy produces a positive correlation between $\theta$ and $\phi$ in the joint posterior, allowing both parameters to shift toward either smaller or larger values. Consistent with this interpretation, AGN for which the inferred viewing angles are substantially smaller than the dynamical modeling estimates generally have large opening angles inferred from dynamical modeling.

In dynamical modeling, $\phi$ is treated as a free parameter. In contrast, we calculate $\phi$ from the FRADO cloud distribution specified by $M$ and $\lambda$ (see \S~\ref{sec:infmodel}). This prescription yields a narrower range of $\phi$ with typically smaller values than those inferred from dynamical modeling. Thus, it is necessary to further study the degeneracy between $\theta$ and $\phi$ to confirm the relatively low viewing angles inferred for the \Hb\ RM sample. We further discuss the opening angle implementation in Section~\ref{sec:openingangle}. Although observationally challenging, spectro-polarimetric constraints on the viewing angle, once available for individual AGN in a larger RM sample, will provide an important independent test of the inferred viewing angle distribution.

\subsection{Representative Viewing Angle $\langle \theta \rangle$}
\label{sec:theta_avg_discussion}

The inferred $\theta$ of individual AGN slightly depends on the choice of $\langle\theta\rangle$, for which we adopt $28.8^\circ$ based on the 37 AGN with dynamical modeling results (see \S~\ref{theta_avg}). As a consistency check, we derive an alternative $\langle\theta\rangle$ assuming an isotropic $\theta$ distribution \citep[see also][]{Lawrence10}. For randomly oriented AGN, the viewing angle probability density, $p(\theta)$, follows $p(\theta)\propto\sin\theta$, where $p(\theta)$ is the probability density per unit viewing angle. Assuming that type~1 AGN are observed within $\theta\leq60^\circ$, the mean of this distribution is $\sim39.2^\circ$. Using this distribution, we define the average log virial factor as
\begin{equation}
\left\langle \log f \right\rangle
=
\frac{
\int_{0^\circ}^{60^\circ}
\log f(\theta)\sin\theta\,d\theta
}{
\int_{0^\circ}^{60^\circ}
\sin\theta\,d\theta
}.
\end{equation}
and determine $\langle \theta \rangle$ from
\begin{equation}
\log f\left(\langle \theta \rangle\right)
=
\left\langle \log f \right\rangle,
\end{equation}
where the viewing angle dependence of the virial factor is expressed as
\begin{equation}
f(\theta)
\propto
\left[
\left(\sin^2\theta+f_{turb}\tan^2\phi\right)
\left(1+\beta\sin\theta\right)
\right]^{-1}.
\end{equation}
This expression follows from the mass calibration in Equation~\ref{eq:M_true}.
From Equations~\ref{eq:openingangle} and \ref{eq:beta}, we obtain approximate ranges of $\phi\sim10^\circ$--$25^\circ$ and $\beta\sim1.3$--$1.7$ over the typical range of $\log(\lambda M)\sim6$--8 in the \Hb\ RM sample. Furthermore, our Bayesian inference constrains $f_{turb}$ to $\sim0.04$--$0.06$ across $\langle\theta\rangle\sim30^\circ$--$40^\circ$. For the adopted ranges of $\phi$, $\beta$, and $f_{turb}$, the corresponding representative viewing angle varies only from $\langle\theta\rangle\sim35.0^\circ$ to $35.5^\circ$, indicating weak sensitivity to these inputs. We therefore adopt $\langle\theta\rangle=35^\circ$ for the alternative configuration.

Using this alternative $\langle\theta\rangle=35^\circ$, we repeat the Bayesian inference and compare the posterior-predictive $\tau_{model}$ with $\tau_{obs}$. We find consistency between $\tau_{model}$ and $\tau_{obs}$, with an offset of $0.041\pm0.22$ dex and an intrinsic scatter of $0.10$ dex.

Next, we compare the inferred $\theta$ for $\langle\theta\rangle=35^\circ$ with the dynamical modeling estimates, as in Section~\ref{sec:theta_discussion}. We obtain an offset of $-2.7^\circ\pm11^\circ$, showing better agreement than for $\langle\theta\rangle=28.8^\circ$. As future \Hb\ RM samples become larger and less affected by selection effects, the representative viewing angle adopted in our Bayesian framework may approach $\langle\theta\rangle=35^\circ$.

Lastly, we revisit the inferred viewing angle distribution of the \Hb\ RM sample for the alternative choice of $\langle\theta\rangle=35^\circ$. The marginalized $\theta$ distribution shifts toward larger viewing angles. Nonetheless, the resulting $\theta$ distribution still shows a substantially larger fraction at small viewing angles than the isotropic distribution.

\subsection{Opening Angle}
\label{sec:openingangle}

\begin{figure*}[!t]
  \centering
  \begin{subfigure}[t]{0.48\textwidth}
   \vspace{0pt}\centering
    \includegraphics[height=0.93\textwidth]{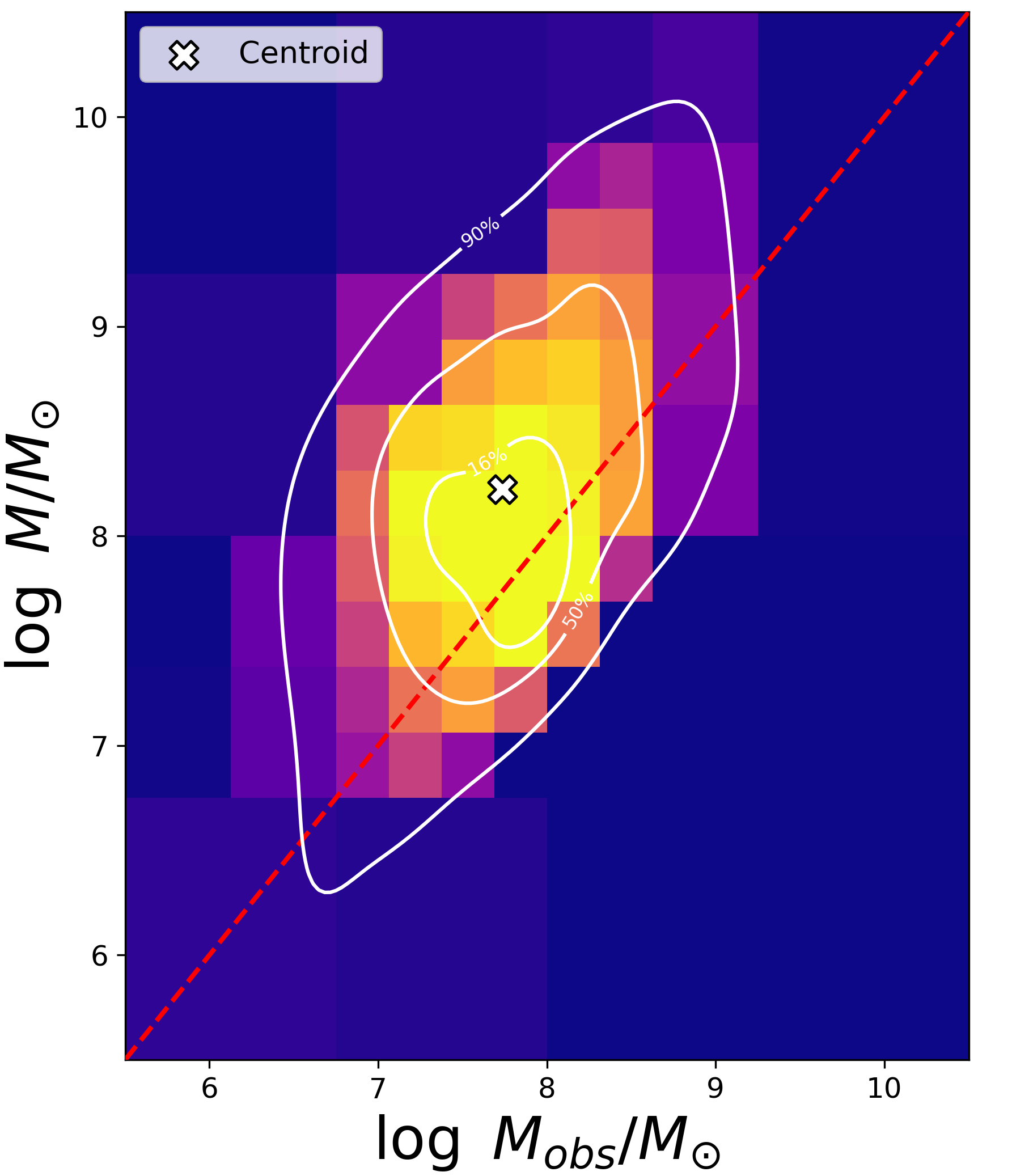}
    \label{REdd&M_comp(a)}
  \end{subfigure}
  \begin{subfigure}[t]{0.48\textwidth}
   \vspace{0pt}\centering
    \includegraphics[height=0.93\textwidth]{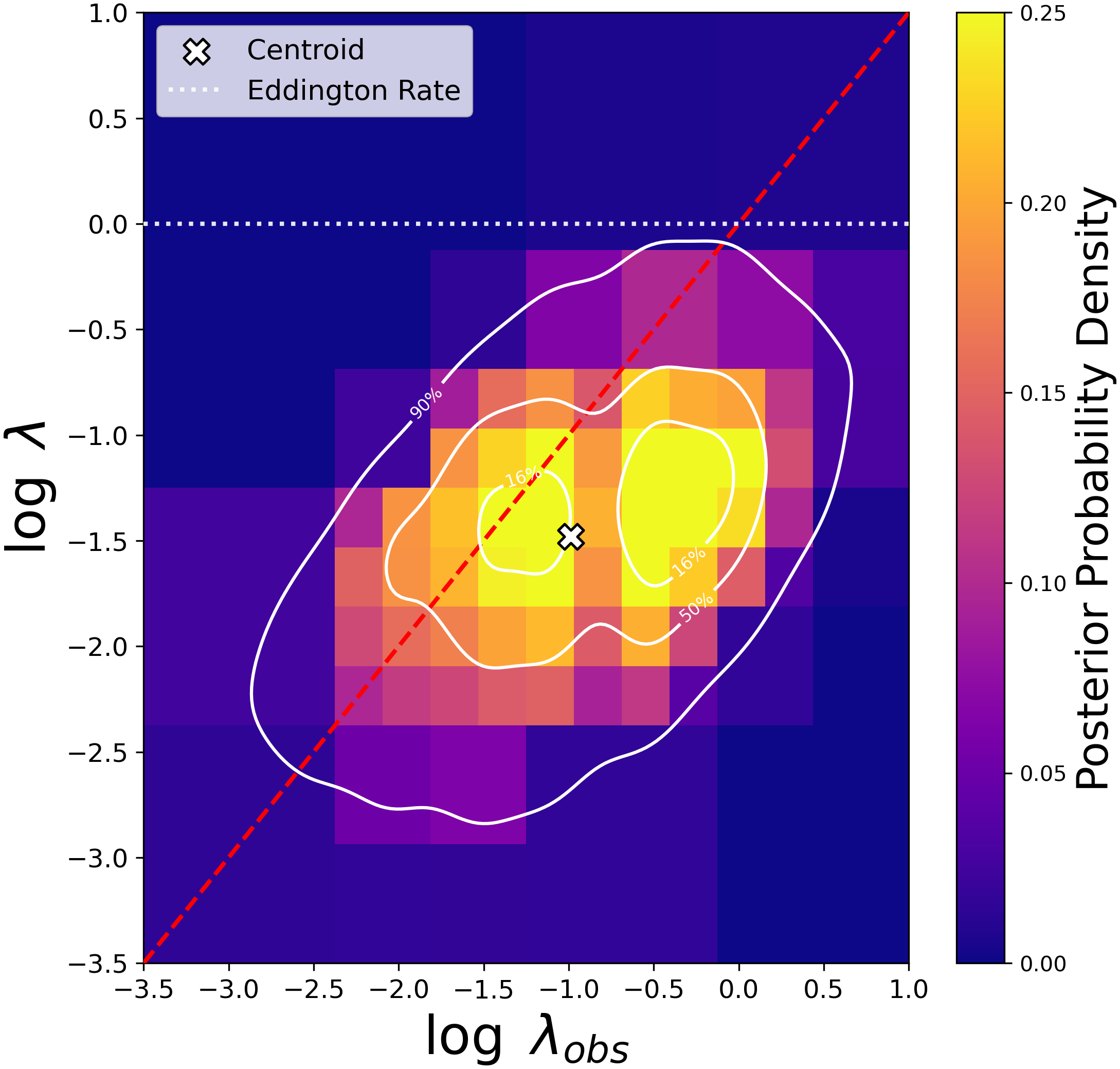}
    \label{REdd&M_comp(b)}
  \end{subfigure}
  \caption{Left: Comparison between the RM \mbh\ ($M_{obs}$) and the intrinsic mass ($M$). Right: Comparison between the observed Eddington ratio ($\lambda_{obs}$) and the intrinsic $\lambda$. White contours denote the 16$\%$, 50$\%$, and 90$\%$ probability regions and the centroids of the distributions are indicated by crosses.}
  \label{REdd&M_comp}
\end{figure*}

In our Bayesian framework, $\phi$ modulates the viewing angle dependence of $\Delta V_{los}^2$ through the turbulent contribution $f_{turb}\tan^2\phi$, where the free parameter $f_{turb}$ controls its amplitude (see \S~\ref{sec:m_lambda_calibration}). The inferred $f_{turb}$ is $\sim0.045$ ($\sim0.056$ for $\langle\theta\rangle=35^\circ$), implying that the opening angle $\phi$ contributes to the turbulent-to-rotational velocity ratio with a coefficient of $\sqrt{f_{turb}}\sim0.21$--$0.24$. Over the typical ranges of $\theta$ and $\phi$ for the \Hb\ RM sample, the rotational component remains dominant, while the turbulent component associated with $\phi$ is subdominant but non-negligible, weakening the viewing angle dependence of the line velocity. This result is qualitatively consistent with \citet{Grier17a}, who reported that the viewing angle has a greater effect on the line profile shape than the opening angle.

As discussed in Section~\ref{sec:theta_discussion}, the degeneracy between $\theta$ and $\phi$ is an important source of uncertainty in the inferred $\theta$. In this work, we calculate $\phi$ from the angular extent of the FRADO cloud distribution (see \S~\ref{sec:m_lambda_calibration}). The resulting opening angles rarely exceed $30^\circ$, even for the most luminous AGN in the \Hb\ RM sample. In contrast, the opening angles inferred from dynamical modeling span approximately $10^\circ$--$70^\circ$ and are typically larger than those predicted by the FRADO model. In both our inference and BLR dynamical modeling, the degeneracy between $\theta$ and $\phi$ produces a positive correlation, such that a larger $\phi$ is associated with a larger $\theta$ (see Section~\ref{sec:theta_discussion}). The limited range of $\phi$ in our model may therefore contribute both to the overall preference for small viewing angles in the \Hb\ RM sample and to the substantially smaller values inferred for some AGN relative to the dynamical modeling estimates. A more realistic treatment of cloud irradiation in future versions of the FRADO model may shift the predicted $\phi$ distribution toward larger values and broaden its range.

\subsection{Systematic Uncertainties of the Observed AGN Properties}
\label{sec:calibration_effects}

The observed RM \mbh\ ($M_{obs}$) and Eddington ratio ($\lambda_{obs}$) deviate from the calibrated mass ($M$) and Eddington ratio ($\lambda$) owing to viewing angle effects. By comparing the observed and intrinsic parameters, we discuss the systematic uncertainties of the observed mass and Eddington ratio. Using the same sampling configuration as in Section~\ref{sec:theta_posterior}, we construct joint probability density maps of $(M_{obs},M)$ and $(\lambda_{obs},\lambda)$ from $4\times10^4$ paired realizations per AGN. To account for measurement uncertainties, we generate 25 Monte Carlo draws of $M_{obs}$ and $\lambda_{obs}$, select one random draw for each realization, and apply the viewing angle calibration using the sampled $\theta$ to obtain $M$ and $\lambda$.

As presented in \figref{REdd&M_comp}, the observed \mbh\ is smaller than $M$ by $0.49\pm0.69$ dex. In contrast, $\lambda_{obs}$ is larger than $\lambda$ by $0.49\pm0.72$ dex. The upward mass shift suggests that the RM \mbh\ of the \Hb\ RM sample may be systematically underestimated, implying that the fraction of super-Eddington AGN may be lower than that estimated using the RM \mbh. In fact, only $\sim1.1\%$ of the calibrated $\lambda$ draws exceed unity, whereas 13 of the 151 AGN ($\sim8.6\%$) have $\lambda_{obs}>1$.

The systematically larger intrinsic \mbh\ appears to be in tension with \citet{Wang26}, who reported that dynamical modeling masses are consistent on average with the RM \mbh. By comparing the dynamical-model-based \mbh\ reported by \citet{Villafana26} with the inferred intrinsic $M$ for the 31 overlapping AGN, we find an offset of $0.49\pm0.38$ dex. The trend toward larger mass in our Bayesian framework relative to the dynamical modeling estimates is more prominent for high-mass AGN. Since the mass discrepancy is driven by the preference for small inferred viewing angles in the \Hb\ RM sample, the degeneracy between $\theta$ and $\phi$ may provide an underlying explanation (see \S~\ref{sec:theta_discussion}). In particular, the limited range of $\phi$ predicted by our model (see also \S~\ref{sec:openingangle}) may restrict the corresponding range of $\theta$, leading to smaller inferred viewing angles and hence higher calibrated masses than those obtained from dynamical modeling.

\section{Summary and Conclusions}
\label{sec:sum&conclusion}
In this work, we present the updated FRADO model and the Bayesian framework to investigate the viewing angle dependence of the \RBLR--\Lmon\ relation.
We summarize our main results as follows:

\begin{itemize}

\item We updated the FRADO model by including a phase factor for anisotropic visibility of the BLR cloud surface to the line-of-sight. We find that the far-side BLR has
a larger weight than the near-side BLR in the transfer function, increasing the viewing angle dependence of the BLR time lag predicted by the model.

\item Using a Bayesian framework, we constrained the viewing angle of each AGN in the \Hb\ RM sample, finding that the inferred viewing angles are relatively small and broadly consistent with the viewing angle estimates based on dynamical modeling of velocity-resolved lags 
%(\citealt{Villafana26}). 
\citep{Villafana26}.
Several objects showed significant offsets, presumably due to the large opening angle inferred from dynamical modeling, reflecting the degeneracy between the viewing angle and the BLR opening angle.

\item 
The updated FRADO combined with a Bayesian framework reproduces the observed \RBLR--\Lmon\ relation. As the viewing angle affects the \Hb\ time lag and the anisotropic disk luminosity, the slope and scatter of the observed \RBLR--\Lmon\ relation can be reproduced with the inferred viewing angles for the \Hb\ RM sample.
%depend on the viewing angle distribution of the AGN sample. 

%%while the true \RBLR--\Lmon\ of the FRADO follows the photoionization prediction. 

\item
Since the RM \mbh\ is determined using an average virial factor corresponding to the representative viewing angle of the sample, the FRADO-based \mbh\ is larger than the RM-based \mbh\ by $\sim0.5$ dex because the viewing angles inferred for individual AGN are smaller than the representative viewing angle.
This systematic mass offset requires further investigation, particularly regarding the limited range of opening angles predicted by the FRADO model.

\end{itemize}

Overall, the updated FRADO model successfully reproduces the observed \RBLR--\Lmon\ relation. 
The dependence of \RBLR, \Lmon, and \mbh\ on viewing angle should be investigated further using both observational and theoretical approaches. 

% while the relatively narrow range of the BLR opening angles constructed by the FRADO model requires further investigation and improvement of the FRADO model.

\begin{acknowledgments}
This work has been supported by the Basic Science Research Program through the National Research Foundation of the Korean Government (RS-2026-25487266). 
\end{acknowledgments}

\clearpage
\bibliography{references_20260720}

@ARTICLE{GRAVITY24,
  author        = {A. Amorim and G. Bourdarot and W. Brandner and Y. Cao and Y. Cl{\'e}net and R. Davies and P. T. de Zeeuw and J. Dexter and A. Drescher and A. Eckart and F. Eisenhauer and M. Fabricius and H. Feuchtgruber and N. M. F{\"o}rster Schreiber and P. J. V. Garcia and R. Genzel and S. Gillessen and D. Gratadour and S. H{\"o}nig and M. Kishimoto and S. Lacour and D. Lutz and F. Millour and H. Netzer and T. Ott and T. Paumard and K. Perraut and G. Perrin and B. M. Peterson and P. O. Petrucci and O. Pfuhl and M. A. Prieto and S. Rabien and D. Rouan and D. J. D. Santos and J. Shangguan and T. Shimizu and A. Sternberg and C. Straubmeier and E. Sturm and L. J. Tacconi and K. R. W. Tristram and F. Widmann and J. Woillez},
  title         = {The Size--Luminosity Relation of Local Active Galactic Nuclei from Interferometric Observations of the Broad-Line Region},
  journal       = {\aap},
  year          = {2024},
  volume        = {684},
  eid           = {A167},
  pages         = {A167},
  doi           = {10.1051/0004-6361/202348167}
}

@ARTICLE{Amorim21,
  author        = {A. Amorim and M. Baub{\"o}ck and W. Brandner and M. Bolzer and Y. Cl{\'e}net and R. Davies and P. T. De Zeeuw and J. Dexter and A. Drescher and A. Eckart and F. Eisenhauer and N. M. F{\"o}rster Schreiber and F. Gao and P. J.V. Garcia and R. Genzel and S. Gillessen and D. Gratadour and S. H{\"o}nig and D. Kaltenbrunner and M. Kishimoto and S. Lacour and D. Lutz and F. Millour and H. Netzer and T. Ott and T. Paumard and K. Perraut and G. Perrin and B. M. Peterson and P. O. Petrucci and O. Pfuhl and M. A. Prieto and D. Rouan and J. Sanchez-Bermudez and J. Shangguan and T. Shimizu and M. Schartmann and J. Stadler and A. Sternberg and O. Straub and C. Straubmeier and E. Sturm and L. J. Tacconi and K. R.W. Tristram and P. Vermot and S. Von Fellenberg and I. Waisberg and F. Widmann and J. Woillez},
  title         = {The central parsec of NGC 3783: a rotating broad emission line region, asymmetric hot dust structure, and compact coronal line region},
  journal       = {\aap},
  year          = {2021},
  volume        = {648},
  eid           = {A117},
  pages         = {A117},
  doi           = {10.1051/0004-6361/202040061}
}

@ARTICLE{Amorim20,
  author        = {A. Amorim and M. Baub{\"o}ck and W. Brandner and Y. Cl{\'e}net and R. Davies and P. T. De Zeeuw and J. Dexter and A. Eckart and F. Eisenhauer and N. M. F{\"o}rster Schreiber and F. Gao and P. J.V. Garcia and R. Genzel and S. Gillessen and D. Gratadour and S. H{\"o}nig and M. Kishimoto and S. Lacour and D. Lutz and F. Millour and H. Netzer and T. Ott and T. Paumard and K. Perraut and G. Perrin and B. M. Peterson and P. O. Petrucci and O. Pfuhl and M. A. Prieto and D. Rouan and J. Shangguan and T. Shimizu and M. Schartmann and J. Stadler and A. Sternberg and O. Straub and C. Straubmeier and E. Sturm and L. J. Tacconi and K. R.W. Tristram and P. Vermot and S. Von Fellenberg and I. Waisberg and F. Widmann and J. Woillez},
  title         = {The spatially resolved broad line region of IRAS 09149-6206},
  journal       = {\aap},
  year          = {2020},
  volume        = {643},
  eid           = {A154},
  pages         = {A154},
  doi           = {10.1051/0004-6361/202039067}
}

@ARTICLE{Sturm18,
  author        = {E. Sturm and J. Dexter and O. Pfuhl and M. R. Stock and R. I. Davies and D. Lutz and Y. Cl{\'e}net and A. Eckart and F. Eisenhauer and R. Genzel and D. Gratadour and S. F. H{\"o}nig and M. Kishimoto and S. Lacour and F. Millour and H. Netzer and G. Perrin and B. M. Peterson and P. O. Petrucci and D. Rouan and I. Waisberg and J. Woillez and A. Amorim and W. Brandner and N. M.F{\"o}rster Schreiber and P. J.V. Garcia and S. Gillessen and T. Ott and T. Paumard and K. Perraut and S. Scheithauer and C. Straubmeier and L. J. Tacconi and F. Widmann},
  title         = {Spatially resolved rotation of the broad-line region of a quasar at sub-parsec scale},
  journal       = {Nature},
  year          = {2018},
  volume        = {563},
  pages         = {657-660},
  doi           = {10.1038/s41586-018-0731-9}
}

@ARTICLE{Shen11,
  author        = {Yue Shen and Gordon T. Richards and Michael A. Strauss and Patrick B. Hall and Donald P. Schneider and Stephanie Snedden and Dmitry Bizyaev and Howard Brewington and Viktor Malanushenko and Elena Malanushenko and Dan Oravetz and Kaike Pan and Audrey Simmons},
  title         = {A Catalog of Quasar Properties from SDSS DR7},
  journal       = {\apjs},
  year          = {2011},
  volume        = {194},
  eid           = {45},
  pages         = {45},
  doi           = {10.1088/0067-0049/194/2/45}
}

@MISC{Pucha26,
  author        = {Ragadeepika Pucha and S. Juneau and M. Mezcua and Arjun Dey and Y. -Y. Mao and D. M. Alexander and C. Circosta and V. A. Fawcett and Wei-Jian Guo and J. Moustakas and S. Panda and M. Siudek and Z. Yu and J. Aguilar and S. Ahlen and D. Bianchi and D. Brooks and T. Claybaugh and K. S. Dawson and A. de la Macorra and P. Doel and S. Ferraro and A. Font-Ribera and J. E. Forero-Romero and E. Gazta{\~n}aga and Satya Gontcho A Gontcho and G. Gutierrez and C. Hahn and K. Honscheid and R. Joyce and R. Kehoe and T. Kisner and A. Kremin and M. Landriau and L. Le Guillou and M. Manera and A. Meisner and R. Miquel and S. Nadathur and W. J. Percival and F. Prada and I. P{\'e}rez-R{\`a}fols and G. Rossi and E. Sanchez and D. Schlegel and M. Schubnell and J. Silber and D. Sprayberry and G. Tarl{\'e} and B. A. Weaver and H. Zou},
  title         = {A New Record Census of Dwarf AGN and a Bimodal $M_{\rm BH}$--$M_{\star}$ Scaling Relation with DESI DR1},
  year          = {2026},
  archivePrefix = {arXiv},
  eprint        = {2606.02699},
  primaryClass  = {astro-ph.GA}
}

@ARTICLE{Pan25,
  author        = {Zhiwei Pan and Linhua Jiang and Wei-Jian Guo and Shengxiu Sun and Ma{\l}gorzata Siudek and Jessica Nicole Aguilar and Steven Ahlen and David Brooks and Todd Claybaugh and Axel de la Macorra and Peter Doel and Enrique Gazta{\~n}aga and Satya Gontcho A Gontcho and Stephanie Juneau and Theodore Kisner and Andrew Lambert and Martin Landriau and Laurent Le Guillou and Marc Manera and Paul Martini and Aaron Meisner and Ramon Miquel and John Moustakas and Adam Myers and Claire Poppett and Francisco Prada and Graziano Rossi and Eusebio Sanchez and Michael Schubnell and Hee-Jong Seo and David Sprayberry and Gregory Tarl{\'e} and Benjamin Alan Weaver and Hu Zou},
  title         = {Iron-corrected Single-epoch Black Hole Masses of DESI Quasars at Low Redshift},
  journal       = {\apj},
  year          = {2025},
  volume        = {987},
  eid           = {48},
  pages         = {48},
  doi           = {10.3847/1538-4357/add7dd}
}

@ARTICLE{WuShen22,
  author        = {Qiaoya Wu and Yue Shen},
  title         = {A Catalog of Quasar Properties from Sloan Digital Sky Survey Data Release 16},
  journal       = {\apjs},
  year          = {2022},
  volume        = {263},
  eid           = {42},
  pages         = {42},
  doi           = {10.3847/1538-4365/ac9ead}
}

@ARTICLE{Rakshit20,
  author        = {S. Rakshit and C. S. Stalin and J. Kotilainen},
  title         = {Spectral Properties of Quasars from Sloan Digital Sky Survey Data Release 14: The Catalog},
  journal       = {\apjs},
  year          = {2020},
  volume        = {249},
  eid           = {17},
  pages         = {17},
  doi           = {10.3847/1538-4365/ab99c5}
}

@ARTICLE{Vestergaard06,
  author        = {Marianne Vestergaard and Bradley M. Peterson},
  title         = {Determining Central Black Hole Masses in Distant Active Galaxies and Quasars. II. Improved Optical and UV Scaling Relationships*},
  journal       = {\apj},
  year          = {2006},
  volume        = {641},
  pages         = {689--711},
  doi           = {10.1086/500572}
}

@ARTICLE{WooUrry02,
  author        = { Jong Hak Woo and C Megan Urry},
  title         = {The Independence of Active Galactic Nucleus Black Hole Mass and Radio Loudness},
  journal       = {\apjl},
  year          = {2002},
  volume        = {581},
  pages         = {L5--L7}
}

@ARTICLE{Harrison24,
  author        = {Chris M. Harrison and Cristina Ramos Almeida},
  title         = {Observational Tests of Active Galactic Nuclei Feedback: An Overview of Approaches and Interpretation},
  journal       = {Galaxies},
  year          = {2024},
  volume        = {12},
  eid           = {17},
  pages         = {17},
  doi           = {10.3390/galaxies12020017}
}

@ARTICLE{Hardcastle20,
  author        = {M. J. Hardcastle and J. H. Croston},
  title         = {Radio galaxies and feedback from AGN jets},
  journal       = {New Astronomy Reviews},
  year          = {2020},
  volume        = {88},
  eid           = {101539},
  pages         = {101539},
  doi           = {10.1016/j.newar.2020.101539}
}

@ARTICLE{Baron19,
  author        = {Dalya Baron and Brice M{\'e}nard},
  title         = {Black Hole Mass Estimation for Active Galactic Nuclei from a New Angle},
  journal       = {\mnras},
  year          = {2019},
  volume        = {487},
  pages         = {3404--3418},
  doi           = {10.1093/mnras/stz1546}
}

@ARTICLE{Peterson14,
  author        = {B. M. Peterson},
  title         = {Measuring the Masses of Supermassive Black Holes},
  journal       = {Space Sci. Rev.},
  year          = {2014},
  volume        = {183},
  pages         = {253--275},
  doi           = {10.1007/s11214-013-9987-4}
}

@ARTICLE{Peterson04,
  author        = {B. M. Peterson and L. Ferrarese and K. M. Gilbert and S. Kaspi and M. A. Malkan and D. Maoz and D. Merritt and H. Netzer and C. A. Onken and R. W. Pogge and M. Vestergaard and A. Wandel},
  title         = {Central Masses and Broad-Line Region Sizes of Active Galactic Nuclei. II. A Homogeneous Analysis of a Large Reverberation-Mapping Database},
  journal       = {\apj},
  year          = {2004},
  volume        = {613},
  pages         = {682--699},
  isbn          = {005452.2+252539},
  doi           = {10.1086/423269}
}

@ARTICLE{Bentz09,
  author        = {Misty C. Bentz and Jonelle L. Walsh and Aaron J. Barth and Nairn Baliber and Vardha Nicola Bennert and Gabriela Canalizo and Alexei V. Filippenko and Mohan Ganeshalingam and Elinor L. Gates and Jenny E. Greene and Marton G. Hidas and Kyle D. Hiner and Nicholas Lee and Weidong Li and Matthew A. Malkan and Takeo Minezaki and Yu Sakata and Frank J.D. Serduke and Jeffrey M. Silverman and Thea N. Steele and Daniel Stern and Rachel A. Street and Carol E. Thornton and Tommaso Treu and Xiaofeng Wang and Jong Hak Woo and Yuzuru Yoshii},
  title         = {The Lick AGN Monitoring Project: Broad-Line Region Radii and Black Hole Masses from Reverberation Mapping of H$eta$},
  journal       = {\apj},
  year          = {2009},
  volume        = {705},
  pages         = {199--217},
  doi           = {10.1088/0004-637X/705/1/199}
}

@ARTICLE{Blandford82,
  author        = {R. D. Blandford and C. F. McKee},
  title         = {Reverberation mapping of the emission line regions of Seyfert galaxies and quasars.},
  journal       = {\apj},
  year          = {1982},
  volume        = {255},
  pages         = {419--439},
  doi           = {10.1086/159843}
}

@ARTICLE{Peterson93,
  author        = {Bradley M. Peterson},
  title         = {Reverberation Mapping of Active Galactic Nuclei},
  journal       = {\pasp},
  year          = {1993},
  volume        = {105},
  pages         = {247--268},
  doi           = {10.1086/133140}
}

@ARTICLE{Du18,
  author        = {Pu Du and Zhi-Xiang Zhang and Kai Wang and Ying-Ke Huang and Yue Zhang and Kai-Xing Lu and Chen Hu and Yan-Rong Li and Jin-Ming Bai and Wei-Hao Bian and Ye-Fei Yuan and Luis C. Ho and Jian-Min Wang and (SEAMBH collaboration)},
  title         = {Supermassive Black Holes with High Accretion Rates in Active Galactic Nuclei. IX. 10 New Observations of Reverberation Mapping and Shortened H$\beta$ Lags},
  journal       = {\apj},
  year          = {2018},
  volume        = {856},
  eid           = {6},
  pages         = {6},
  doi           = {10.3847/1538-4357/AAAE6B}
}

@ARTICLE{Naddaf21,
  author        = {Mohammad Hassan Naddaf and Bo{\.z}ena Czerny and Ryszard Szczerba},
  title         = {The Picture of BLR in 2.5D FRADO: Dynamics and Geometry},
  journal       = {\apj},
  year          = {2021},
  volume        = {920},
  eid           = {30},
  pages         = {30},
  doi           = {10.3847/1538-4357/ac139d}
}

@ARTICLE{Naddaf20,
  author        = {Mohammad Hassan Naddaf and Bo{\.z}ena Czerny and Ryszard Szczerba},
  title         = {BLR Size in Realistic FRADO Model: The Role of Shielding Effect},
  journal       = {Front. Astron. Space Sci.},
  year          = {2020},
  volume        = {7},
  eid           = {15},
  pages         = {15},
  doi           = {10.3389/fspas.2020.00015}
}

@ARTICLE{Czerny17,
  author        = {B. Czerny and Yan-Rong Li and K. Hryniewicz and S. Panda and C. Wildy and M. Sniegowska and J.-M. Wang and J. Sredzinska and V. Karas},
  title         = {Failed Radiatively Accelerated Dusty Outflow Model of the Broad Line Region in Active Galactic Nuclei. I. Analytical Solution},
  journal       = {\apj},
  year          = {2017},
  volume        = {846},
  eid           = {154},
  pages         = {154},
  doi           = {10.3847/1538-4357/AA8810}
}

@ARTICLE{Naddaf25,
  author        = {M. H. Naddaf and M. L. Martinez-Aldama and D. Hutsemekers and D. Savic and B. Czerny},
  title         = {H$eta$ Line Shape and Radius--Luminosity Relation in 2.5D FRADO},
  journal       = {\aap},
  year          = {2025},
  volume        = {702},
  eid           = {A46},
  pages         = {A46},
  doi           = {10.1051/0004-6361/202555767}
}

@ARTICLE{Czerny11,
  author        = {B. Czerny and K. Hryniewicz},
  title         = {The origin of the broad line region in active galactic nuclei},
  journal       = {\aap},
  year          = {2011},
  volume        = {525},
  eid           = {L8},
  pages         = {L8},
  doi           = {10.1051/0004-6361/201016025}
}

@article{Pancoast14,
   author = {Anna Pancoast and Brendon J. Brewer and Tommaso Treu and Daeseong Park and Aaron J. Barth and Misty C. Bentz and Jong Hak Woo},
   doi = {10.1093/MNRAS/STU1419},
   issn = {0035-8711},
   issue = {3},
   journal = {Monthly Notices of the Royal Astronomical Society},
   month = {12},
   pages = {3073-3091},
   publisher = {Oxford Academic},
   title = {Modelling reverberation mapping data – II. Dynamical modelling of the Lick AGN Monitoring Project 2008 data set},
   volume = {445},
   url = {https://dx.doi.org/10.1093/mnras/stu1419},
   year = {2014}
}

@article{Peterson11,
   author = {Bradley M Peterson},
   month = {9},
   title = {Masses of Black Holes in Active Galactic Nuclei: Implications for NLS1s},
   url = {https://arxiv.org/pdf/1109.4181},
   year = {2011}
}

@ARTICLE{Sbarrato18,
  author        = {Tullia Sbarrato and Massimo Dotti and Giancarlo Ghirlanda and Fabrizio Tavecchio},
  title         = {On the Orientation of Narrow-Line Seyfert 1 Galaxies},
  journal       = {\aap},
  year          = {2018},
  volume        = {616},
  eid           = {A43},
  pages         = {A43},
  doi           = {10.1051/0004-6361/201732451}
}

@INPROCEEDINGS{Nagao01,
  author        = {T. Nagao and Y. Taniguchi and T. Murayama},
  title         = {How Do We See Narrow-Line Seyfert 1 Galaxies? Properties of Emission-Line Regions in Narrow-Line Seyfert 1 Galaxies},
  booktitle     = {Narrow-Line Seyfert 1 Galaxies and Their Place in the Universe},
  series        = {Astronomical Society of the Pacific Conference Series},
  year          = {2001},
  volume        = {222},
  pages         = {379},
  isbn          = {1-58381-054-4}
}

@INPROCEEDINGS{Murayama01,
  author        = {T. Murayama and Y. Taniguchi and T. Nagao},
  title         = {A New Pole-on View Model for Narrow-Line Seyfert 1 Galaxies},
  booktitle     = {Narrow-Line Seyfert 1 Galaxies and Their Place in the Universe},
  series        = {Astronomical Society of the Pacific Conference Series},
  year          = {2001},
  volume        = {222},
  pages         = {375},
  isbn          = {1-58381-054-4}
}

@ARTICLE{Bian04,
  author        = {W. Bian and Y. Zhao},
  title         = {Black hole masses in narrow-line Seyfert 1 galaxies},
  journal       = {\mnras},
  year          = {2004},
  volume        = {352},
  pages         = {823--827},
  doi           = {10.1111/j.1365-2966.2004.07967.x}
}

@ARTICLE{Decarli08,
  author        = {Roberto Decarli and Massimo Dotti and Marcella Fontana and Francesco Haardt},
  title         = {Are the black hole masses in narrow-line Seyfert 1 galaxies actually small?},
  journal       = {\mnras},
  year          = {2008},
  volume        = {386},
  pages         = {L15--L19},
  doi           = {10.1111/j.1745-3933.2008.00451.x}
}

@ARTICLE{Pancoast12,
  author        = {{Pancoast}, Anna and {Brewer}, Brendon J. and {Treu}, Tommaso and {Barth}, Aaron J. and {Bennert}, Vardha N. and {Canalizo}, Gabriela and {Filippenko}, Alexei V. and {Gates}, Elinor L. and {Greene}, Jenny E. and {Li}, Weidong and {Malkan}, Matthew A. and {Sand}, David J. and {Stern}, Daniel and {Woo},  Jong Hak and {Assef}, Roberto J. and {Bae}, Hyun-Jin and {Buehler}, Tabitha and {Cenko}, S. Bradley and {Clubb}, Kelsey I. and {Cooper}, Michael C. and {Diamond-Stanic}, Aleksandar M. and {Hiner}, Kyle D. and {H{\"o}nig}, Sebastian F. and {Joner}, Michael D. and {Kandrashoff}, Michael T. and {Laney}, C. David and {Lazarova}, Mariana S. and {Nierenberg}, A.~M. and {Park}, Dawoo and {Silverman}, Jeffrey M. and {Son}, Donghoon and {Sonnenfeld}, Alessandro and {Thorman}, Shawn J. and {Tollerud}, Erik J. and {Walsh}, Jonelle L. and {Walters}, Richard},
  title         = {{The Lick AGN Monitoring Project 2011: Dynamical Modeling of the Broad-line Region in Mrk 50}},
  journal       = {\apj},
  year          = {2012},
  volume        = {754},
  number        = {1},
  eid           = {49},
  pages         = {49},
  doi           = {10.1088/0004-637X/754/1/49},
  archivePrefix = {arXiv},
  eprint        = {1205.3789},
  primaryClass  = {astro-ph.CO}
}

@ARTICLE{Li21,
  author        = {Sha Sha Li and Sen Yang and Zi Xu Yang and Yong Jie Chen and Yu Yang Songsheng and He-Zhen Liu and Pu Du and Bin Luo and Zhe Yu and Chen Hu and Bo Wei Jiang and Dong-Wei Bao and Wei-Jian Guo and Zhi-Xiang Zhang and Yan-Rong Li and Ming Xiao and Kai-Xing Lu and Luis C. Ho and Jin-Ming Bai and Wei-Hao Bian and Jes{\'u}s Aceituno and Takeo Minezaki and Keith Horne and Mitsuru Kokubo and Jian-Min Wang},
  title         = {Reverberation Mapping of Two Luminous Quasars: The Broad-line Region Structure and Black Hole Mass},
  journal       = {\apj},
  year          = {2021},
  volume        = {920},
  eid           = {9},
  pages         = {9},
  doi           = {10.3847/1538-4357/AC116E}
}

@ARTICLE{Du16,
  author        = {Pu Du and Kai-Xing Lu and Zhi-Xiang Zhang and Ying-Ke Huang and Kai Wang and Chen Hu and Jie Qiu and Yan-Rong Li and Xu-Liang Fan and Xiang-Er Fang and Jin-Ming Bai and Wei-Hao Bian and Ye-Fei Yuan and Luis C. Ho and Jian-Min Wang and (SEAMBH collaboration)},
  title         = {SUPERMASSIVE BLACK HOLES WITH HIGH ACCRETION RATES IN ACTIVE GALACTIC NUCLEI. V. A NEW SIZE--LUMINOSITY SCALING RELATION FOR THE BROAD-LINE REGION},
  journal       = {\apj},
  year          = {2016},
  volume        = {825},
  eid           = {126},
  pages         = {126},
  doi           = {10.3847/0004-637X/825/2/126}
}

@ARTICLE{Yu19,
  author        = {Li Ming Yu and Bi Xuan Zhao and Wei Hao Bian and Chan Wang and Xue Ge},
  title         = {An Extended Size--Luminosity Relation for the Reverberation-Mapped AGNs: The Role of the Accretion Rate},
  journal       = {\mnras},
  year          = {2020},
  volume        = {491},
  pages         = {5881--5896},
  doi           = {10.1093/mnras/stz3387}
}

@ARTICLE{Du19,
  author        = {Pu Du and Jian Min Wang},
  title         = {The Radius-Luminosity Relationship Depends on Optical Spectra in Active Galactic Nuclei},
  journal       = {\apj},
  year          = {2019},
  volume        = {886},
  eid           = {42},
  pages         = {42},
  doi           = {10.3847/1538-4357/ab4908}
}

@ARTICLE{Wang24,
  author        = {Shu Wang and Jong Hak Woo},
  title         = {Revisiting the H$\beta$ Size--Luminosity Relation Using a Uniform Reverberation-mapping Analysis},
  journal       = {\apjs},
  year          = {2024},
  volume        = {275},
  eid           = {13},
  pages         = {13},
  doi           = {10.3847/1538-4365/AD74F2}
}

@ARTICLE{Wang26,
  author        = {Shu Wang and Jong Hak Woo and Lizvette Villafa{\~n}a and Tommaso Treu and Elena Gallo},
  title         = {Dynamical Modeling of the Broad-line Region with High-mass Active Galactic Nuclei and Constraints on the Virial Factor},
  journal       = {\apj},
  year          = {2026},
  volume        = {1000},
  eid           = {180},
  pages         = {180},
  doi           = {10.3847/1538-4357/AE4699}
}

@ARTICLE{Villafana26,
  author        = {Lizvette Villafa{\~n}a and Tommaso Treu and Shu Wang and Misty C Bentz and Brendon J Brewer and Aaron J Barth and  Jong Hak Woo and Matthew A Malkan and Vardha N Bennert},
  title         = {Can Broad-line Region Line Profile Shape Improve Single-epoch Black Hole Mass Estimates?},
  journal       = {\apj},
  year          = {2026},
  volume        = {1004},
  eid           = {151},
  pages         = {151},
  doi           = {10.3847/1538-4357/ae69cf}
}

@ARTICLE{Woo26,
  author        = { Jong Hak Woo and Jimin Kim and Hojin Cho and Shu Wang},
  title         = {New Black Hole Mass Calibrations and the Fundamental Plane of the Broad-line Region Size, Luminosity, and Velocity},
  journal       = {\apj},
  year          = {2026},
  volume        = {1003},
  eid           = {123},
  pages         = {123},
  doi           = {10.3847/1538-4357/ae6109}
}

@ARTICLE{Pancoast11,
  author        = {Anna Pancoast and Brendon J Brewer and Tommaso Treu},
  title         = {Geometric and Dynamical Models of Reverberation Mapping Data},
  journal       = {\apj},
  year          = {2011},
  volume        = {730},
  eid           = {139},
  pages         = {139},
  doi           = {10.1088/0004-637X/730/2/139}
}

@ARTICLE{Williams18,
  author        = {Peter R. Williams and Anna Pancoast and Tommaso Treu and Brendon J. Brewer and Aaron J. Barth and Vardha N. Bennert and Tabitha Buehler and Gabriela Canalizo and S. Bradley Cenko and Kelsey I. Clubb and Michael C. Cooper and Alexei V. Filippenko and Elinor Gates and Sebastian F. Hoenig and Michael D. Joner and Michael T. Kandrashoff and Clifton David Laney and Mariana S. Lazarova and Weidong Li and Matthew A. Malkan and Jacob Rex and Jeffrey M. Silverman and Erik Tollerud and Jonelle L. Walsh and  Jong Hak Woo},
  title         = {The Lick AGN Monitoring Project 2011: Dynamical Modeling of the Broad-line Region},
  journal       = {\apj},
  year          = {2018},
  volume        = {866},
  eid           = {75},
  pages         = {75},
  doi           = {10.3847/1538-4357/AAE086}
}

@ARTICLE{Lawrence10,
  author        = {A. Lawrence and M. Elvis},
  title         = {Misaligned Disks as Obscurers in Active Galaxies},
  journal       = {\apj},
  year          = {2010},
  volume        = {714},
  pages         = {561--570},
  doi           = {10.1088/0004-637X/714/1/561}
}

@ARTICLE{Grier17a,
  author        = {C. J. Grier and A. Pancoast and A. J. Barth and M. M. Fausnaugh and B. J. Brewer and T. Treu and B. M. Peterson},
  title         = {The Structure of the Broad-line Region in Active Galactic Nuclei. II. Dynamical Modeling of Data From the AGN10 Reverberation Mapping Campaign},
  journal       = {\apj},
  year          = {2017},
  volume        = {849},
  eid           = {146},
  pages         = {146},
  doi           = {10.3847/1538-4357/aa901b}
}

@ARTICLE{Collin06,
  author        = {{Collin}, S. and {Kawaguchi}, T. and {Peterson}, B.~M. and {Vestergaard}, M.},
  title         = {{Systematic effects in measurement of black hole masses by emission-line reverberation of active galactic nuclei: Eddington ratio and inclination}},
  journal       = {\aap},
  year          = {2006},
  volume        = {456},
  number        = {1},
  pages         = {75--90},
  doi           = {10.1051/0004-6361:20064878},
  archivePrefix = {arXiv},
  eprint        = {astro-ph/0603460},
  primaryClass  = {astro-ph}
}

@ARTICLE{Bentz13,
  author        = {Misty C Bentz and Kelly D Denney and Catherine J Grier and Aaron J Barth and Bradley M Peterson and Marianne Vestergaard and Vardha N Bennert and Gabriela Canalizo and Gisella De Rosa and Alexei V Filippenko and Elinor L Gates and Jenny E Greene and Weidong Li and Matthew A Malkan and Richard W Pogge and Daniel Stern and Tommaso Treu and  Jong Hak Woo},
  title         = {THE LOW-LUMINOSITY END OF THE RADIUS-LUMINOSITY RELATIONSHIP FOR ACTIVE GALACTIC NUCLEI},
  journal       = {\apj},
  year          = {2013},
  volume        = {767},
  eid           = {149},
  pages         = {149},
  doi           = {10.1088/0004-637X/767/2/149}
}

@ARTICLE{Kaspi00,
  author        = {Shai Kaspi and Paul S. Smith and Hagai Netzer and Dan Maoz and Buell T. Jannuzi and Uriel Giveon},
  title         = {Reverberation Measurements for 17 Quasars and the Size-Mass-Luminosity Relations in Active Galactic Nuclei},
  journal       = {\apj},
  year          = {2000},
  volume        = {533},
  pages         = {631--649},
  doi           = {10.1086/308704}
}

@ARTICLE{Heckman14,
  author        = {Timothy M. Heckman and Philip N. Best},
  title         = {The coevolution of galaxies and supermassive black holes: Insights from surveys of the contemporary universe},
  journal       = {\araa},
  year          = {2014},
  volume        = {52},
  pages         = {589--660},
  doi           = {10.1146/annurev-astro-081913-035722}
}

@ARTICLE{Fabian12,
  author        = {A. C. Fabian},
  title         = {Observational Evidence of AGN Feedback},
  journal       = {\araa},
  year          = {2012},
  volume        = {50},
  pages         = {455--489},
  doi           = {10.1146/ANNUREV-ASTRO-081811-125521}
}

@ARTICLE{Jaiswal25,
  author        = {Vikram Kumar Jaiswal and Amit Kumar Mandal and Raj Prince and Ashwani Pandey and Mohammad Hassan Naddaf and Bo{\.z}ena Czerny and Swayamtrupta Panda and Francisco Pozo Nu{\~n}ez},
  title         = {Application of the FRADO model of broad line region formation to Seyfert galaxy NGC 5548 and a first step toward determining the Hubble constant},
  journal       = {\aap},
  year          = {2025},
  volume        = {702},
  eid           = {A92},
  pages         = {A92},
  doi           = {10.1051/0004-6361/202452497}
}

@ARTICLE{Naddaf24,
  author        = {{Naddaf}, Mohammad Hassan},
  title         = {{The Effect of Outflow Launching Radial Efficiency of Accretion Disk on the Shape of Emission-Line Profiles}},
  journal       = {Universe},
  year          = {2024},
  volume        = {11},
  number        = {1},
  eid           = {8},
  pages         = {8},
  doi           = {10.3390/universe11010008},
  archivePrefix = {arXiv},
  eprint        = {2412.18772},
  primaryClass  = {astro-ph.GA}
}

@ARTICLE{Murray95,
  author        = {N. Murray and J. Chiang and S. A. Grossman and G. M. Voit},
  title         = {Accretion Disk Winds from Active Galactic Nuclei},
  journal       = {\apj},
  year          = {1995},
  volume        = {451},
  pages         = {498--509},
  doi           = {10.1086/176238}
}

@ARTICLE{Woo24,
  author        = {Jong Hak Woo and Shu Wang and Suvendu Rakshit and Hojin Cho and Donghoon Son and Vardha N. Bennert and Elena Gallo and Edmund Hodges Kluck and Tommaso Treu and Aaron J. Barth and Wanjin Cho and Adi Foord and Jaehyuk Geum and Hengxiao Guo and Yashashree Jadhav and Yiseul Jeon and Kyle M. Kabasares and Won Suk Kang and Changseok Kim and Minjin Kim and Tae Woo Kim and Huynh Anh N. Le and Matthew A. Malkan and Amit Kumar Mandal and Daeseong Park and Chance Spencer and Jaejin Shin and Hyun il Sung and Vivian U and Peter R. Williams and Nick Yee},
  title         = {The Seoul National University AGN Monitoring Project. III. H$\beta$ Lag Measurements of 32 Luminous Active Galactic Nuclei and the High luminosity End of the Size--Luminosity Relation},
  journal       = {\apj},
  year          = {2024},
  volume        = {962},
  eid           = {67},
  pages         = {67},
  doi           = {10.3847/1538-4357/AD132F}
}

@ARTICLE{Martinez20,
  author        = {Mary Loli Mart{\'i}nez--Aldama and Michal Zaja{\v{c}}ek and Bo{\.z}ena Czerny and Swayamtrupta Panda},
  title         = {Scatter Analysis along the Multidimensional Radius--Luminosity Relations for Reverberation-mapped Mg ii Sources},
  journal       = {\apj},
  year          = {2020},
  volume        = {903},
  eid           = {86},
  pages         = {86},
  doi           = {10.3847/1538-4357/abb6f8}
}

@article{Ferland89,
   author = {G J Ferland and S E Persson},
   journal = {The Astrophysical Journal},
   pages = {656-673},
   title = {IMPLICATIONS OF Ca n EMISSION FOR PHYSICAL CONDITIONS IN THE BROAD-LINE REGION OF ACTIVE GALACTIC NUCLEI},
   volume = {347},
   year = {1989}
}

@article{Emmering92,
   author = {Robert T. Emmering and Roger D. Blandford and Isaac Shlosman and Robert T. Emmering and Roger D. Blandford and Isaac Shlosman and Emmering and Robert T. and Blandford and Roger D. and Shlosman and Isaac},
   doi = {10.1086/170955},
   issn = {0004-637X},
   journal = {ApJ},
   month = {2},
   pages = {460},
   publisher = {American Astronomical Society},
   title = {Magnetic Acceleration of Broad Emission-Line Clouds in Active Galactic Nuclei},
   volume = {385},
   url = {https://ui.adsabs.harvard.edu/abs/1992ApJ...385..460E/abstract},
   year = {1992}
}

@article{Begelman17,
   author = {Mitchell C. Begelman and Joseph Silk},
   doi = {10.1093/MNRAS/STW2533},
   issn = {0035-8711},
   issue = {2},
   journal = {Monthly Notices of the Royal Astronomical Society},
   month = {1},
   pages = {2311-2317},
   publisher = {Oxford Academic},
   title = {Magnetically elevated accretion discs in active galactic nuclei: broad emission-line regions and associated star formation},
   volume = {464},
   url = {https://dx.doi.org/10.1093/mnras/stw2533},
   year = {2017}
}

@article{Baskin18,
   author = {Alexei Baskin and Ari Laor},
   doi = {10.1093/MNRAS/STX2850},
   issn = {0035-8711},
   issue = {2},
   journal = {Monthly Notices of the Royal Astronomical Society},
   month = {2},
   pages = {1970-1994},
   publisher = {Oxford Academic},
   title = {Dust inflated accretion disc as the origin of the broad line region in active galactic nuclei},
   volume = {474},
   url = {https://dx.doi.org/10.1093/mnras/stx2850},
   year = {2018}
}

@article{Netzer93,
   author = {Netzer, Hagai and Laor, Ari},
   doi = {10.1086/186741},
   issn = {0004-637X},
   journal = {ApJL},
   month = {2},
   pages = {L51},
   publisher = {American Astronomical Society},
   title = {Dust in the Narrow-Line Region of Active Galactic Nuclei},
   volume = {404},
   url = {https://ui.adsabs.harvard.edu/abs/1993ApJ...404L..51N/abstract},
   year = {1993}
}

@article{Hopkins25,
   author = {Philip F. Hopkins},
   doi = {10.33232/001c.137969},
   issn = {25656120},
   journal = {Open Journal of Astrophysics},
   pages = {56},
   publisher = {National University of Ireland Maynooth},
   title = {Multi-phase thermal structure \& the origin of the broad-line region, torus, and corona in magnetically-dominated accretion disks},
   volume = {8},
   url = {https://ui.adsabs.harvard.edu/abs/2025OJAp....8E..56H/abstract},
   year = {2025}
}

@article{Woo15,
   author = {Jong Hak Woo and Yosep Yoon and Songyoun Park and Daeseong Park and Sang Chul Kim},
   doi = {10.1088/0004-637X/801/1/38},
   issn = {0004-637X},
   issue = {1},
   journal = {The Astrophysical Journal},
   month = {2},
   pages = {38},
   publisher = {IOP Publishing},
   title = {THE BLACK HOLE MASS–STELLAR VELOCITY DISPERSION RELATION OF NARROW-LINE SEYFERT 1 GALAXIES},
   volume = {801},
   url = {https://iopscience.iop.org/article/10.1088/0004-637X/801/1/38 https://iopscience.iop.org/article/10.1088/0004-637X/801/1/38/meta},
   year = {2015}
}
\bibliographystyle{aasjournalv7}

\end{document}